\documentclass[a4paper,12pt,twoside]{report}
\usepackage[left=2cm,right=2cm,top=2cm,bottom=3cm]{geometry}

\usepackage{graphicx}
\usepackage{verbatim}
\usepackage{latexsym}
\usepackage{mathchars}
\usepackage{setspace}
\input{blocked.sty}
\input{uhead.sty}
\input{boxit.sty}
\input{icthesis.sty}
\usepackage{subfig}

\newcommand{\normallinespacing}{\renewcommand{\baselinestretch}{1.5} \normalsize}

\newcommand{\syncc}{~\stackrel{\textstyle \rhd\kern-0.57em\lhd}{\scriptstyle L}~}

\usepackage{mathtools}
\usepackage{amssymb}
\usepackage{amsmath}
\usepackage[colorlinks=true, allcolors=blue]{hyperref}

\graphicspath{{images/}}

\DeclarePairedDelimiter\floor{\lfloor}{\rfloor}
\usepackage{listings}
\usepackage{color}
\definecolor{dkgreen}{rgb}{0,0.6,0}
\definecolor{dkorange}{rgb}{0.8,0.3,0}
\definecolor{gray}{rgb}{0.5,0.5,0.5}
\definecolor{mauve}{rgb}{0.58,0,0.82}

\lstdefinestyle{customsh}{
  belowcaptionskip=1\baselineskip,
  breaklines=true,
  frame=L,
  xleftmargin=\parindent,
  showstringspaces=false,
  basicstyle=\footnotesize\ttfamily,
  keywordstyle=\color{dkgreen},
  commentstyle=\color{dkorange},
  breaklines=true,
  breakatwhitespace=true,
}

\lstdefinestyle{customc}{
  belowcaptionskip=1\baselineskip,
  breaklines=true,
  frame=L,
  xleftmargin=\parindent,
  showstringspaces=false,
  basicstyle=\footnotesize\ttfamily,
  keywordstyle=\color{dkgreen},
  morecomment=[l][{\color{dkorange}}]{\#},
  commentstyle=\color{red},
  stringstyle=\color{mauve},
  identifierstyle=\color{blue},
  breaklines=true,
  breakatwhitespace=true,
}

\begin{document}

\title{\LARGE {\bf Applying the Polyhedral Model to Tile Loops in Devito}\\
 \vspace*{6mm}
}

\author{Dylan McCormick}
\submitdate{June 2017}

\normallinespacing
\maketitle

\addcontentsline{toc}{chapter}{Abstract}

\begin{abstract}

The run time of many scientific computation applications for numerical methods
is heavily dependent on just a few multi-dimensional loop nests. Since these applications are 
often limited by memory bandwidth rather than computational resources they can benefit 
greatly from any optimizations which decrease the run time of their loops by
improving data reuse and thus reducing the total memory traffic. Some of the most
effective of these optimizations are not suitable for development by hand
or require advanced software engineering knowledge which 
is beyond the level of many researchers who are not specialists in code optimization. 
Several tools exist to automate the generation of high-performance code for numerical methods,
such as Devito which produces code for finite-difference approximations typically used
in the seismic imaging domain. We present a loop-tiling optimization which can be applied
to Devito-generated loops and improves run time by up to 27.5\%, and options for automating 
this optimization in the Devito framework.

\end{abstract}

\cleardoublepage

\addcontentsline{toc}{chapter}{Acknowledgements}

\begin{acknowledgements}

I would like to express gratitude to my supervisor Fabio Luporini for his enthusiastic support,
excellent advice and friendliness throughout this project, which he managed to give despite his own sizeable responsibilities. 
The guidance, expertise and teachings of Prof. Paul H J Kelly were invaluable to my taking interest
in the domain of high-performance code and computer architecture, and even more so to my completion of this project.
I certainly give my thanks as well to Sun Tianjiao who provided vital insight at several stages of the project
and was always perpared to help in our regular meetings.

Finally, I owe this work and much more to my parents for believing in me and giving me the opportunity to be here. 

\end{acknowledgements}

\body
\chapter{Introduction} 
There are many established optimizations which improve the run-time performance of program loops and have been in use for decades.
The theory of why these optimizations work has been researched extensively \cite{uniform-loop} and has been practically applied in modern optimizing compilers so
transformed code need not always be hand-written. Some transformations (such as loop unrolling or tiling) 
can have a negative impact on the readability and maintainability of code (even if performed manually by a developer)
by increasing code size and changing statement ordering. This can make them undesirable or even infeasible to develop and maintain by hand, in spite
of their performance benefits. It is an attractive option to have a smart compiler or code generator make the transformations automatically so the
original computation description remains concise. 

Automatically identifying and taking advantage of opportunities for loop optimizations is a tremendous
challenge for an optimizing tool which has to ensure program correctness based on the information available in the source code. If the tool cannot guarantee
correctness or appreciate the logical implications of a piece of code then no transformation occurs, resulting in a potential performance opportunity cost.
Some tools operate in a more domain-specific context where the problems they handle pertain to a particular field; for example, seismic imaging.
Such tools have more knowledge about possible inputs and what requirements applications in the domain have. This allows them to rely on more
assumptions and often apply more aggressive optimizations.


Automated code generation is an increasingly important solution to the problem of writing high-performance code
for scientific applications. Decades of research into compiler optimizations and advanced computer architecture developments
have uncovered a number of techniques that can be applied in translating high-level problem descriptions into low-level source code
to achieve some form of performance increase. The complexity of modern systems and computational problems makes it 
difficult for a software engineer to produce maximally efficient code by hand even with the help of smart
compilers. This is more challenging for scientists and researchers whose specialization is not in high-performance software engineering. Tools
and frameworks for automatic generation of application-specific high-performance code from high-level or even symbolic input are a very
attractive option for improving performance of numerical computations. Devito is one such tool that generates high-performance code from symbolic
input which approximates partial differential equations typically applied in the seismic imaging domain. It currently applies a number 
of optimizations and analyses \cite{devito-fd} when generating code, but could benefit from additional loop optimizations that improve performance 
for some problems by increasing data reuse. Loop bodies are typically where scientific computing applications spend most of their time \cite{dragon}, and when dealing with
real-world large-scale problems run times can easily be on the order of hours, so every relative improvement in run time can have significant
benefits for users.

\section{Loop tiling}
One optimization that can improve the performance of a loop nest is known as `loop tiling' or `loop blocking', which improves data locality and parallelism
by transforming the way a loop nests iterates over its domain. Loop tiling is one of the primary optimizations researched during this project.
Loops which have been tiled can sometimes be further optimized in ways that were previously impossible. For example, 
loop-invariant code motion (LICM) transformations on non-tiled code can be restricted by available memory when `moving'
statements from within a loop with a high number of iterations. This limitation can be overcome when LICM is applied to tiled code. Such
loop nests and transformations lend themselves well to geometric reasoning as loops can be considered as dimensions in space and time and individual
iterations as integral points within those dimensions, as in the polyhedral model \cite{poly_framework} . This gives rise to a host of theory and tools (so-called polyhedral tools) which apply polyhedral 
techniques to reason about and optimize loop nests.
\clearpage

\section{Contributions}
The contributions of this project are as follows;
\begin{itemize}
	\item{A demonstration of loop-tiling optimizations that can be applied to space-time loop nests in Devito stencils using 
    the polyhedral model. (\autoref{ch:timetiling})}
    \item{Several time-tiled Devito stencil source files and the CLooG inputs used to generate their loops. \cite{my_code}}
	\item{An evaluation of the functional correctness and runtime performance of the generated code. (\autoref{ch:evaluation})}
    \item{Evidence that legal time-tiling transformations can reduce the run time of some Devito stencil loops by up to 
    27.5\% (\autoref{sec:results})}
	\item{A discussion of potential implementations for automating time-tiling transformations in Devito to 
    increase run-time performance gains realized by end users. (\autoref{ch:conclusion})}
\end{itemize}

\section{Report Structure}
This report describes the research and methods used to investigate the tiling of time loops
in Devito stencil codes. To better communicate the motivation and findings of this research, we 
first define some of the core concepts, tools and techniques that factored into our work in \autoref{ch:background}. 
With that understanding we introduce Devito, the code-generation framework in which the proposed optimizations 
are to operate. In \autoref{ch:tiling} we build on the fundamentals in \autoref{ch:background} and define the loop
transformations which are used to perform the time-tiling optimization described in \autoref{ch:timetiling}.
We present our methods for evaluating the correctness and performance of time-tiled code as well as
results from a number of experiments in \autoref{ch:evaluation}. To conclude, we summarise our investigation
and discuss possible choices for future work in automating time-tiling in Devito.
\chapter{Background}
\label{ch:background}

The study of automated loop optimizations brings together knowledge from compiler design, software engineering and computer architecture
in the pursuit of generating high-performance code for a given application in a reliable manner. In the following sections
we describe some of the core concepts in these fields and use this to motivate a discussion of three tools currently available 
for code generation and transformation in the polyhedral model; CLooG, PLuTo and Pochoir.

\section{Fundamentals}
In this section we introduce some of the fundamental concepts and terminology used throughout this research.
In particular we cover the basics of compilers, loops, stencils, data dependencies and data locality.
\subsection{Compilers}
A compiler is a piece of software which produces code in a target language by assembling and interpreting input code in a source language.
Typically, compilers are used to translate a high-level programming language which describes computation to a human developer (like C or Java) into 
a low-level programming language (like assembly) which is understood by a computer. Due to the complexity of modern languages, systems and computer
architectures this is not a straightforward translation and compilers have a lot of freedom in how they choose to interpret, transform and analyse
their input in order to produce output that meets certain requirements. For example, a compiler might reorder certain instructions from the input
program to produce output code which executes faster than without the instruction reordering.

In addition to common `general purpose compilers' (such as \texttt{gcc} or \texttt{javac} which compile C and Java programs respectively),
there are `high-level compilers' which don't generate low-level executable code and `domain-specific compilers' which sometimes operate on
custom language inputs and are designed for programs in a particular field.

\subsection{Loops}
A common construct in modern code, especially when implementing numerical methods, is the `loop'. The majority of programs spend most of their
time executing loops \cite{dragon}, so compiler designers who are interested in optimization dedicate much research to 
loop transformations and analyses in an effort to reduce their performance impact. The type of loop dealt with in this project is the
`\texttt{for} loop' which is typically structured as in \ref{fig:forloop}. A \texttt{for} loop has a `body' of statements which are repeatedly executed
while the `loop condition' holds. There are no special constraints on the types of statements that can feature in the body of a \texttt{for} loop so it
is possible and indeed very common to have a \texttt{for} loop present in the body of another \texttt{for} loop. Placement of loops within other loops in such a 
is called `nesting' and gives rise to the `loop nest' construct which refers to a group of nested loops. When a loop nest is structured
such that all of its statements are found in the innermost loop body it is known as a `perfect loop nest'.

\begin{figure}[h]
\centering
\begin{lstlisting}[language=C, style=customc]
  //Loop header defines bounds and step operation
  for(int i=0; i < LIMIT; i++){
      //Loop body contains code executed at each step
      ...
  }
  \end{lstlisting}
  \caption{The structure of a simple \texttt{for} loop in C}
  \label{fig:forloop}
\end{figure}

\subsection{Iteration spaces}
When analysing and describing iterative computations, particularly with nested iterative components, it can be useful to consider 
all of the iterations described by the code as points in an `iteration space' \cite{dragon}. Each point in an iteration space represents a
particular assignment of values to the loop indices. Iteration spaces provide a geometric description of loop nests which can be
transformed and visualised. By considering connections between points in an iteration space, the order of iteration or 
inter-iteration dependencies can also be described. Iteration spaces are used throughout this paper to explain the research and results obtained.

Take this simple code fragment which iterates over a two-dimensional array and sets its elements to 0
\begin{figure}[h]
\centering
\begin{lstlisting}[language=C, style=customc]
  int array2d[5][10];
  
  for(int i=0; i<5; i++){
      for(int j=0; j<10; j++){
          array2d[i][j] = 0;
      }
  }
\end{lstlisting}
    \caption{
        This is loop nest iterates over a two-dimensional structure and sets all of its elements to 0. It is an example
        of a `perfect loop nest'.
    }
    \label{fig:2dloop}
\end{figure}

Its iteration space is a rectangle of points representing all the combinations of the loop-indices that are within bounds.

\begin{figure}[h]
\centering
    \centering
    \includegraphics[scale=0.85]{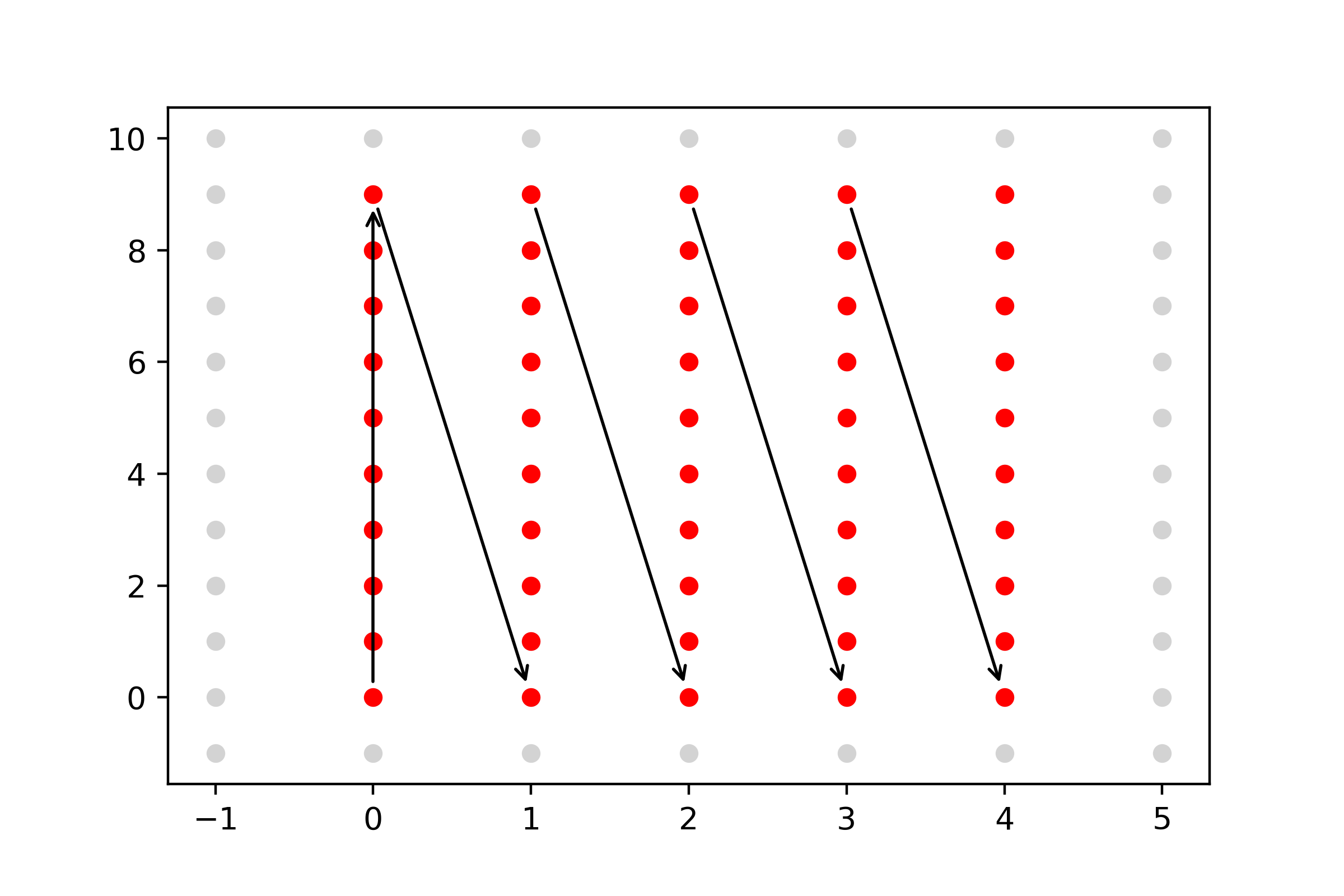}
    \caption{A representation of the iteration space for \ref{fig:2dloop} where the horizontal and vertical axes 
    correspond to loops \texttt{i} and \texttt{j} respectively. Arrows indicate the direction of iteration.}
    \label{fig:iteration_space}
\end{figure}

A more complex two-dimensional loop nest might perform a computation at each iteration which requires a value from an earlier
computation thereby creating a dependence of on previous iterations. In \ref{fig:2dloop} the computation is trivial and there are no
inter-iteration dependencies since the loop body makes no reference to array elements calculated in earlier iterations.

\begin{figure}[h]
\centering
\begin{lstlisting}[language=C, style=customc]
  int array2d[100][100];

  for(int i=1; i<100; i++){
      for(int j=1; j<100; j++){
          array2d[i][j] = array2d[i-1][j] 
                        + 2*array2d[i][j-1];
      }
  }
\end{lstlisting}
\caption{
    This loop nest iterates over a two-dimensional structure and updates the value of all but the first elements in each dimension.
        The right hand of the assignment refers to two array indices which were calculated in previous iterations of the \texttt{i} and \texttt{j} loops
        respectively. It is an example of a `perfect loop nest' as well as a `stencil loop'.
    }
    \label{fig:stencil}
\end{figure}

\subsection{Stencils}
A computation which iterates over a grid and performs nearest-neighbour computations is known as a stencil . These are
particularly common when implementing code that deals with numerical methods and partial differential equations.
Each point in the grid is updated some weighted contributions from a subset of its neighbours \cite{stencil}. \ref{fig:stencil} is an example of a stencil.

\subsection{Data Dependencies}
When any kind of loop transformations are being performed it is important that the loop's dependence structure in understood
to ensure that dependencies are not violated. A data dependence occurs between two statements
if they both reference the same memory and at least one of the statements writes to that memory \cite{dragon}. The loop
transformations presented in this paper do not modify statements in a loop body, so rather than reasoning
about dependencies at the statement level, it is more helpful to consider how dependencies form between loop
iterations. Dependencies between different iterations of the same loop are called \textit{loop-carried} dependencies
in contrast to \textit{loop-dependent} dependencies which occur between statements in a single iteration.
Depending on the original ordering of the statements there are three fundamental types of dependency;
\begin{itemize}
\item \textbf{True dependence}: An assignment is used in a subsequent iteration (read-after-write). Also known as a `flow dependence'.
\item \textbf{Anti dependence}: A location read in one iteration is overwritten in a subsequent iteration (write-after-read).
\item \textbf{Output dependence}: An assignment is overridden by a later iteration. (write-after-write).
\end{itemize}

Dependency violation can occur any time statements (or iterations) are executed in a different order than the
order in which they were written. This is particularly relevant for tiling transformations which group iteration points
together into tiles and iterate over those tiles according to some order defined by the transformation, resulting in
significant changes to the ordering of individual iterations.

\subsubsection{Dependence distance vectors}
A useful tool in understanding loop dependencies, particularly within the polyhedral model, is the 
\textit{dependence distance vector}. Consider the following representation $L^n$ of the set containing all
iteration points of a loop nest with depth $n$;
\[ L^n(\,\vec{I}\,) = \{(I_1,I_2,...,I_n): l_k \leq I_k < U_k \, \forall k.\, 1 \leq k \leq n \}\]
This representation assigns to each iteration of a loop nest a vector whose components are the values
of the loop iterator at the corresponding depth. For example, at the beginning of the 250th iteration of the loop
in \ref{fig:stencil}, the \texttt{i} and \texttt{j} iterators will be 3 and 50 respectively so the vector representation would be
\[ \vec{I} = (I_1,I_2) = (3, 50) \]
If a dependence exists between statements at iterations $\vec{I}_1$ and $\vec{I}_2$, then the \textit{dependence distance}
vector $\vec{d}$ is given by 
\[ \vec{d} = (\vec{I}_2 - \vec{I}_1) \]
which measures the `distance' (number of iterations) in each dimension between the two statements creating the dependency.
By reasoning about this distance, we can determine what (if any) transformations are required to make a tiling legal.

\subsection{Data locality}
Modern computer architectures have a hierarchy of data caches which store data under locality assumptions that memory locations close to each other
are likely to be used within a short period of time (spatial locality) and that a memory location is likely to be reused (temporal locality) \cite{dragon}. 
When statements in a loop body access a memory location, it is loaded into the data cache hierarchy along with nearby memory locations to pre-empt probable future uses.
However non-optimized loop structures may not make use of this data before it is evicted from the cache by another memory access.
A common goal of loop optimizations is to improve data locality across iterations so that expensive fetches from slower memory can be avoided 
and caching techniques are more efficiently utilized.

\section{Loop optimizations}
Loops have a dominant impact on the performance of many programs that make use of them, particularly if loops are nested.
Typically, loop optimizations reshape the
way a loop nest traverses its iteration space to try and exploit \textit{data locality} in the underlying computer architecture which
manages the data used by the loop body. There are several common loop optimizations which transform a loop nest structure i.e) 
exchanging the order of loop headers to change the order of iteration, however the primary type of optimization considered in this 
research is \textit{loop tiling}.

\subsection{Polyhedral model}
The Polyhedral model is a powerful abstraction for reasoning about loop nests and loop transformations \cite{pluto} that builds on the
geometric representation of loop iterations described in an iteration space. Each iteration of a statement is viewed as an
integral point within a polyhedron that contains all iterations of the statement. With a polyhedron (or union of polyhedra) 
for each statement and an understanding
of the dependencies between statements, linear algebra and linear programming techniques can be applied to transform and scan
the iteration spaces they represent.

\section{CLooG}
\label{sec:cloog}
The Polyhedral model is certainly useful for performing transformations on abstract representations of a loop nest,
but it does not deal directly with code. Instead, a separate process is required to turn polyhedral specifications into code
which matches those specifications. In this section we give background information and motivate our usage
of the code generation tool CLooG.

CLooG (Chunky Loop Generator) \cite{cloog} is a free software which generates loops that visit each integral
point found within a convex polyhedron in lexicographical order. It is also designed to be
the code generation back-end to tools that perform automatic parallelism and locality optimizations. 
Its output is pseudo-code loops which visit
each integral point of a union of polyhedra in such a way as to minimize control overhead and produce efficient code.

\subsection{Motivation}
The polyhedral model \cite{ram-polyhedral} allows reasoning about and solving a wide range of problems related to program transformations,
in particular where loop nests are involved. When performing these transformations, code generation is usually the last
step after the abstract structure of the program has been analysed and mutated. There are often constraints on the size of the code
that can be generated to ensure readability for developers and practicality for users attempting to run the code on their machine,
however the most concise code is not always the most efficient, and some transformations may generate complicated loop bounds in an effort
to reduce code size. This type of code can be difficult for a compiler to optimize and for a CPU to schedule in an optimal way due to
poor control management. CLooG provides an interface for applying polyhedral reasoning techniques to generate code which is optimized for control
and within the user's requirements of size or complexity.

We considered CLooG as a candidate for the tiled code generation back-end for Devito because of its simplicity in use and its incorporation into
other tools performing similar tasks as this project's goal. The ability to manually build an input file for a simple example of Devito's
output provided an excellent starting point for understanding how both tools work and for obtaining initial results.

\subsection{Implementation}
CLooG provides a command-line interface which makes use of specially formatted input files (primarily composed of matrices representing
the sets of inequalities defining an iteration space) to understand the loop nest structure as well as a C library that has structures and
functions that permit programmatic configuration and execution of the tool.
Although CLooG is frequently used to make loop-nests parallelizable or more control efficient, it is not concerned with the nature of the
code found within a loop (apart from other loops), and makes no assumptions whatsoever about dependencies between statements.
Statements in CLooG are represented abstractly by inequalities on the iterators which define upon which iterations the statements are executed.
For example the loop body in \ref{fig:stencil} is described to CLooG by its domain, given by the inequalities
\[i \ge 1 \, \wedge \, i < 100 \, \wedge \, j \ge 1 \, \wedge  \, j < 100 \]

\subsubsection{Algorithm}
The algorithm employed by CLooG for visiting the integral points of the specified polyhedra \cite{polyhedra-algo} is presented in in a simplified form.
For each dimension in the iteration space described by the user input, proceeding from outermost to innermost loops;
\begin{enumerate}
    \item Project the polyhedra onto the corresponding dimension 
    \item Separate the projections into disjoint polyhedra which represent, in this dimension, either the overlapped iteration
        space of multiple statements, or the unique iteration space of a single statement. At this stage, the constraints in the current
        dimension give loop bounds that scan the polyhedra in this dimension. The constraints in the other dimensions are guards on 
        the statements in other dimensions.
    \item Recursively repeat this process for each polyhedron, projecting in the next dimension.
    \item Sort the resulting loops so that they scan the polyhedra in lexicographical order.
\end{enumerate}
At this stage CLooG has enough information to `pretty-print' the loops as compilable C code.

\subsection{Limitations}
CLooG is a tool usually used as a back-end to a more specialised tool such as PLUTO \cite{pluto} or PrimeTile \cite{primetile} which
apply polyhedral theory to transform a given loop nest to enhance parallelism or data locality and delegate CLooG to generate
efficient code to scan the new polyhedron. CLooG itself however, is not automated in any way and must be driven with a problem
specification that includes inequalities defining the iteration space, and optional `scattering functions' which encode the scheduling
of statements. This makes standalone usage of CLooG somewhat restrictive, but integration into Devito's code generation pipeline
very attractive. An additional technical constraint of CLooG is that it provides a command-line interface and C libraries only,
limiting its ability to interface with other languages such as Python which is common in scientific computing applications.

\section{PLuTo}
To best understand how loop optimizations can be automated, and how code generation can be driven from abstract code
representation, it is worth considering existing automated loop optimization tools. In this section we discuss the
a tool called PLuTo, its implementation and why it was not used for this research.

PLuTo \cite{pluto} is an open-source tool for automatic polyhedral parallelization and locality optimization. Using CLooG as a code generation
back-end, PLuTo applies analytical model-driven transformations in the polyhedral model \cite{ram-polyhedral} to improve the performance
of regular programs containing loop nests. One of the primary features of PLuTo is its automatic exploration of the space
of potential program transformations to identify and apply the most effective loop tiling.

\subsection{Motivation}
Like CLooG, PLuTo is motivated by the powerful abstraction provided by the Polyhedral model when reasoning about and optimizing loop
nests. However PLuTo is more specifically concerned with how loop tiling can improve parallelism and data locality, especially
for applications run on massively parallel architectures like those frequently seen in scientific computing applications. The framework
takes a step further yet by applying analytical methods to intelligently find optimal transformations and tilings with code dependencies in mind.
The tool's design is indicative of its intended usage in optimizing existing code, possibly developed before a tool as capable as PLuTo was developed.

\subsection{Implementation}
PLuTo operates as a source-to-source optimization tool, taking as input a sequence of nested loops in source-code format
and producing a transformed sequences of loops in source code. The research and behind PLuTo is on automated transformations,
not dependence analysis or source-code parsing, so it delegates the task of interpreting input and determining dependencies
to another tool in the polyhedral space, LooPo. The transformation framework implemented in PLuTo takes polyhedral domains representing the iteration
spaces of the original program, as well as dependence polyhedra which encode the dependence structure of the input source-code.

The transformation framework within PLuTo makes use of a bounding cost function to reason about dependences and iteration domains
geometrically and to provide a target to minimise when identifying optimal transformations. The cost function can be formulated
as a system of Integer Linear Programming (ILP) problems which encode legality and cost constraints and can be solved by the Simplex
algorithm. The framework iteratively augments and solves ILP problems to find multiple independent solutions for each statement. With
these solutions, PLuTo is able to construct matrices representing the transformed domains and statement dependencies which can be
given as input to CLooG for code generation.

\subsection{Limitations}
PLuTo is a powerful tool that makes many informed analyses to drive its automated search for optimal transformations, and its integration
with other polyhedral tools make it a well-documented practical tool for users looking to apply source-to-source loop transformations. It
delivers a more specialised, automated solution than using CLooG alone, and allows with existing code to perform optimizations without having to 
analyse the loop and dependence structure of their program. However
PLuTo's generality and source-to-source nature make it unsuitable for direct application in the Devito pipeline. With the specific structure and
nature of code generated by Devito, assumptions can be made that greatly reduce the work a tool like PLuTo would need to do in determining
dependencies and optimal transformations, but the general approach PLuTo takes prevents these assumptions from being made. The source-code level
is also a low-level representation that restricts what can be done by Devito after the transformations take place and would require premature
generation of source code to feed as input to PLuTo.

\section{Pochoir}
The scientific codes this research is concerned with are stencil loops, which are a specific type of computation that have
common structures, limitations and optimization opportunities. In this section we provide some background on a stencil-oriented
code generation and optimization tool called Pochoir \cite{pochoir}.

Pochoir is a tool for implementing cache-efficient stencil codes on multi-core processors written in C++. It provides a 
domain-specific language embedded in C++ so developers can write functional problem specifications and automatically generate
highly optimized, cache-efficient parallel code. A novel feature of Pochoir is its decomposition of the user workflow
into two steps; safety-checking compilation with Pochoir libraries and the Intel C++ compiler followed by re-compilation with
the Pochoir compiler, producing optimized parallel code.

\subsection{Motivation}
Stencil computations are often conceptually simple to develop using the intuitive single loop nest implementation
that follows directly from the problem specification. However, such implementations are rarely cache-efficient and 
producing more performant, parallelized codes manually can be a significant challenge for a programmer. Pochoir aims
to provide a solution to this challenge which offers a simple user workflow and automatically produces highly optimized
code using advanced polyhedral techniques.

\subsection{Implementation}
Pochoir comprises two main components, the Pochoir Template Library and the Pochoir compiler, which provide
the means to execute and optimize code written in the Pochoir DSL. The template library includes loop-based
and trapezoidal algorithms for executing problem specifications written in the DSL and is used in both steps of
the Pochoir workflow to generate code. In the first step, the template library allows users to execute their Pochoir specification
after compiling with the Intel Compiler \cite{icc}. This provides assurances of correctness and validity before the optimization
stage as users can debug using standard toolchains and ensure that their specification is numerically correct. Following
this, the source is re-compiled using the Pochoir compiler which performs optimizations at the DSL level before finally 
compiling with the Intel compiler and the Template library to produce optimized, parallel code. Pochoir uses 
the Intel CilkPlus framework \cite{icc} for data and task parallelism in its generated code.

The efficiency and parallelism offered by Pochoir is derived from its divide-and-conquer algorithm for recursively
decomposing a loop's iteration space into trapezoidal sections. The algorithm is relatively simple, but relies on
complex results and lemmas regarding the legality of its transformations. To describe it simply, the algorithm 
applies \textit{hyperspace cuts} and \textit{time cuts} to the iteration space, which partition it in the spatial and time
dimensions respectively, and then recursively applies the same procedure to the resulting partitions. By repeatedly analyzing
how dependencies interact with the partitioned iteration spaces, the algorithm ensures that its cuts are valid. The number
and placement of hyperspace and time cuts is chosen in a novel way which aims to maximize opportunities for parallelism.
Following this recursive decomposition, a number of optimizations are applied which simplify and improve
the efficiency of boundary condition code and loop indexing, before finally compiling the transformed specification using
the Intel compiler and the Pochoir Template Library to produce efficient, parallel code.

\subsection{Limitations}
The Pochoir compiler offers a novel approach to domain-specific languages embedded in C++ coupled with 
an advanced process for creating a highly-parallel iteration space and optimized code to iterate through it.
However, Pochoir is coupled to C++, the Intel compiler and the Cilk parallelism framework which make it a powerful
tool for users working with Intel hardware in that environment, but limits its applicability to computing
applications that use other languages, hardware and threading models. The Pochoir DSL is general and expressive,
but is embedded in C++ which can be a challenging language for scientific work by researchers who do not specialise
in high-performance, object-oriented development. The Template Library's use of C++ template metaprogramming 
can also inhibit extension or modification of its algorithms. Additionally, Pochoir's optimizations do not include
code transformations to exploit advanced vector capabilities of the underlying hardware, unlike Devito.

\section{Summary}
We have discussed some of the basic concepts which apply to optimizing scientific codes and introduce
terminology which wil be used throughout this report to talk about stencil codes, iteration spaces, data dependencies
and more. We give a brief evaluation of three polyhedral tools which are related to the work we present here; CLooG
(the primary tool we used to generate our own optimized code), PLuTo and Pochoir. With an understanding of these fundamentals
and polyhedral tools we will next introduce Devito, the framework in which this work is meant to be applied.

\chapter{Devito}
\label{ch:devito}
In this chapter we discuss the purpose, implementation and operation of Devito, the code-generation tool
for which the optimizations evaluated in this research are intended. We outline the context in which Devito
is used, describe its core algorithms, components and optimizations before comparing it to other tools in the same domain.

Devito \cite{devito-fd}, \cite{devito-dsl}, is a domain-specific language (DSL) and code generation framework that produces highly-optimised
code for finite-difference computations \cite{opesci-fd}. It uses high-level symbolic problem descriptions that can be hand-written to automatically
generate and optimise high-performance parallel C code for a range of computer architectures. The finite-difference kernels that
Devito generates are used to solve partial differential equations that are primarily used for simulations typical in the oil and gas industry.

\section{Motivation}
Modern high-performance computing applications need to make use of the latest developments in computer architecture to produce truly
powerful solutions. As modern technologies become more powerful they also become more complex with massively parallel systems like
GPGPU (General Purpose Graphics Processing Unit) and advanced architectural capabilities such as vectorization providing opportunities
to improve program performance at the cost of increased implementation complexity. For scientists looking to leverage efficient 
technologies in their programs the Python programming language is a common choice because of its natural form of expression,
large collection of open-source packages that allow interaction with some of these advanced architectures and its ease of use when compared with traditional languages like C.
However, as an interpreted language Python is not suited for direct use in HPC applications, and currently available methods that reduce the interpreter overhead introduce
additional complexity and are still not as efficient as straight C code. 

Devito aims to combine the ease and expressiveness of Python, with the 
power and optimizations available to C code. By using a domain-specific language embedded in a python package, Devito can provide a specialized interface
that describes numerical computations and perform code generation with aggressive optimizations to produce faster programs that require less HPC expertise
from the developer.

\section{Design}
The high-level process architecture of Devito can be divided into five primary parts;
\begin{enumerate}
    \item Create Devito data objects which associate SymPy function symbols with user data
    \item Build symbolic stencil equations using the created data objects
    \item Build Devito Operator object using symbolic equations
    \item Instruct the Devito Operator to generate low-level optimized code applied to user data
    \item Compile using user-defined compiler settings
\end{enumerate}

\begin{figure}[h]
\centering
  \begin{lstlisting}[language=Python, style=customc]
  from devito import TimeData, Operator
  from sympy.abc import s, h
  # Function symbol associated with user data
  u = TimeData(name="u" shape=(nx, ny),
               time_order=1, space_order=2)
  u.data[0, :] = ui[:]

  # Symbolic equation
  eqn = Eq(u.dt, a * (u.dx2 + u.dy2))
  # Expand finite-difference stencils
  stencil = solve(eqn, u.forward)[0]

  # Devito operator that transforms stencils into array accesses
  op = Operator(stencils=Eq(u.forward, stencil),
                subs={h: dx, s: dt}, nt=timesteps)

  # Generate low-level code applied to data with Propagator
  op.apply()
  \end{lstlisting}
  \caption{ Python set-up code for generating a Diffusion kernel in Devito \cite{devito-dsl}}
  \label{fig:diffusion_devito}
\end{figure}

One of the key benefits of using Devito is the automatic generation of abstract array accesses upon creation of the Operator object. This 
creates an intermediate representation (IR) of the stencil which is then optimized and translated into C when triggered by the user using the Operator
in a process called `propagation'. It is during this propagation stage that loops are generated, and so would be the point where
Devito could analyse the IR and delegate to a polyhedral tool to tile the structure being generated. The objects that make up the IR allow
Devito to programmatically reason about the loop and dependence structure of the kernel which would form the primary input for polyhedral transformations.

\begin{figure}[h]
\centering
\begin{lstlisting}[language=C, style=customc]
  extern "C" int Operator(float *u_vec)
  {
      float (*u)[1000][1000] = (float (*)[1000][1000]) u_vec;
      {
          int t0;
          int t1;
          #pragma omp parallel
          for (int i3 = 0; i3<500; i3+=1)
          {
              #pragma omp single
              {
                  t0 = (i3)%(2);
                  t1 = (t0 + 1)%(2);
              }
              {
                  #pragma omp for schedule(static)
                  for (int i1 = 1; i1<999; i1++)
                  {
                      #pragma omp simd aligned(u:64)
                      for (int i2 = 1; i2<999; i2++)
                      {
                          u[t1][i1][i2] = 2.5e-1F*u[t0][i1][i2-1]
                                        + 2.5e-1F*u[t0][i1][i2+1]
                                        + 2.5e-1F*u[t0][i1-1][i2]
                                        + 2.5e-1F*u[t0][i1+1][i2];
                      }
                  }
              }
          }
      }
  }
\end{lstlisting}
\caption{Auto-generated C code to solve the diffusion example in \ref{fig:diffusion_devito} \cite{devito-dsl}.}
\label{fig:diffusion_C}
\end{figure}

Devito takes its form as a Python library that provides an API for users to create and manage objects that define and configure all of the components
in the code generation pipeline. To make the most of the native Python environment and the practical benefits it brings, Devito does not define its 
own high-level DSL in the traditional sense, instead it makes use of and extends the powerful SymPy Python library for symbolic mathematics. 
The familiar abstraction provided by a native python library allows users to separate implementation and optimization concerns from the numerical computation at hand.

\section{Devito Loop Engine}
The Devito Loop Engine (DLE) is a component of Devito which is responsible for applying loop optimizations
and code transformations. At present, it is structured as a collection of core components for analysing and
modifying ASTs with a number of `backends' that specify custom optimizing transformations running on top.
For both the time-tiled and standard Devito stencil codes evaluated in this paper, a number of optimizations 
provided by the DLE were enabled.

\subsubsection{Denormals}
\textit{Denormal} or \textit{subnormal} numbers are floating point numbers which are smaller than those typically
allowed by the IEEE floating point specification \cite{denormals}. These numbers have reduced precision compared to `normal' numbers 
as the exponent used to represent denormals is already at its minimum, and so smaller values are achieved by truncating
significant digits. While denormals facilitate gradual loss of precision rather than the usual total loss of precision,
they can have an impact on program performance. Devito provides an optimizing pass which `avoids denormals' by instructing
the CPU to set denormal numbers to zero upon encountering them.

\subsubsection{Parallelism}
For large-scale stencil loops such as those generated by Devito, parallelism can offer invaluable performance improvements.
When instructing Devito to generate code using OpenMP parallelism directives \cite{openmp}, a transforming pass inserts two
OMP \textit{pragma} statements around each parallelizable loop nest. First, a \texttt{\#pragma omp parallel} directive
defines the region of code which will be executed in parallel. Then a \texttt{\#pragma omp for schedule(static)}
directive is placed directly above the outermost parallelizable loop which ensures that threads are used to execute
different loop iterations in parallel according to a `static' scheduling system. Upon reaching the \texttt{for} loop,
each thread is statically assigned the set of iterations it should execute.

\subsubsection{Vectorization}
Another type of parallelism which Devito exploits is at the data level through the use of special
\textit{single instruction multiple data} (SIMD) instructions \cite{vectorization}. These instructions are also called vector
instructions as they perform operations simultaneously on multiple elements within a block (vector) of data 
but only require the CPU to process a single instruction. After an analysis pass has determined which
loops can be vectorized (typically the innermost stencil loops) the vectorizing transformation pass decorates
the loop header with \texttt{\#pragma omp ivdep} and \texttt{\#pragma omp simd} directives. These directives
instruct the compiler to ignore vector dependencies in the vectorized block (which is valid as dependencies are carried 
through time in Devito, and only the innermost space dimension is vectorized) and to use SIMD operations
to execute multiple iterations of the loop simultaneously. 

\begin{figure}[h]
\centering
\begin{lstlisting}[language=C, style=customc]
  for (int time = 1; time < time_size - 1; time += 1)
  {
    #pragma omp parallel
    {
      _MM_SET_DENORMALS_ZERO_MODE(_MM_DENORMALS_ZERO_ON);
      _MM_SET_FLUSH_ZERO_MODE(_MM_FLUSH_ZERO_ON);
      #pragma omp for schedule(static)
      for (int x = 4; x < x_size - 4; x += 1)
      {
        for (int y = 4; y < y_size - 4; y += 1)
        {
          #pragma ivdep
          #pragma omp simd
          for (int z = 4; z < z_size - 4; z += 1)
          {
          	//Stencil code
          }
        }
      }
    }
\end{lstlisting}
\caption{A simplified excerpt from a stencil code generated by Devito which demonstrates the usage of instructions
and directives for parallelism, vectorization and avoiding denormals.}
\end{figure}

\section{Summary}
Devito offers powerful abstractions and optimizations to users looking to perform high-performance, parallelized numerical methods 
calculations for finite-difference approximations without needing to be high-performance computing specialists.
Its developers, the OPESCI group, have active users in research and industry and are continue to develop the framework with 
new features and improvements. We have outlined some of the design and implementation details of Devito to give an understanding
of the context in which our research into time-tiling transformations operates. We use this and the previous chapter to motivate
the introduction of our own time-tiling transformation and how it can be applied in Devito, but first a discussion of more 
advanced tiling transformations is required to understand and validate our propositions.

\chapter{Tiling in the Polyhedral Model} 
\label{ch:tiling}
In this chapter we discuss three well-known loop transformations to give important background
information for understanding time-tiling. We build upon the fundamental concepts
from the introduction and introduce notation from the polyhedral model
to formally describe loop tiling, skewing and skewed tiling. 

When tiling loops, there are a number of factors that affect the structure and efficiency 
of the resulting code. Choosing a good tile configuration is crucial to producing code
with the desired improvements in locality or parallelism. Ultimately the parameters chosen 
when producing a transformation result in different dimensions, orientations
and quantities of the tiles. This is represented at the code level by the structure
and bounds of a loop nest or loop nest(s). 

Transforming code with tiling has a direct influence on program performance 
both in terms of run time and memory utilization, as it changes the manner in which data is accessed or computed. 
For example, bounds in tiled loop nests often make use of `min' and `max' functions to
account for partial tiles or feature more intensive calculations than are usually found in
loop bounds, and so a tiling transformation that results in a large number of very small 
tiles can suffer from the higher relative control costs incurred by repeatedly
executing loops with small trip counts. A worst-case transformation violates
the data dependencies of the original code and results in logically different, and probably incorrect, behaviour.

\section{Polyhedral Representation}
The Polyhedral model provides powerful abstractions for formally reasoning about loops and iteration spaces.
By treating iterations as integral lattice points within convex polyhedra, the model allows for generic mathematical
expression of loop transformations and their composition. The framework lends itself well to efficient code
generation and was a vital tool for researching and performing our optimizations.
For the purposes of this research, and to better relate to the input data consumed by CLooG,
we will work with a simplified interpretation of the framework presented in \cite{poly_framework}
In this section we introduce the formal notation and definitions for these concepts.

\subsection{Statement}
In the polyhedral model the statements of interest are those found in the body of a loop nest.
A code-level program statement in a loop body is represented as a tuple $S$ comprising the statement's 
iteration domain  $D^S$ and schedule $\theta^S$ such that $S = (D^S, \theta^S)$. This structure contains all the 
information about \textit{where} in the iteration space space (domain) and \textit{when} in exeuction time (schedule) 
a statement is executed.

\subsection{Statement Domains}
A statement's iteration domain is a \textit{convex polyhedron} described by a matrix $D^S$ of 
\textit{affine inequalities} \cite{poly_tutorial}. A statement's domain is closely related to its \textit{instance set}; the
set of `dynamic execution instances', or the set of executed instances of an abstracted piece of code (such as
a statement which references the value of an iterator) in a 
loop body. The structure of $D^S$ as a matrix varies between implementations so for clarity we will use set 
notation to describe these matrices of inequalities. $D^S$ as a matrix is described in more detail in 
the next chapter in the context of CLooG's input.

\begin{figure}[h]
\centering
\begin{lstlisting}[language=C, style=customc]
for (int i = 0; i < i_ub; t++) {
	for (int j = 0; j < j_ub; x++){
    	S(i,j); //S1
    }
}
\end{lstlisting}
\caption{A simple two-dimensional loop \texttt{S(i, j)} represents an arbitrary statement which 
depends upon the values of \texttt{i} and \texttt{j}.} 
\label{fig:domain_snippet}
\end{figure}

In \ref{fig:domain_snippet}, the domain $D^{S1}$ for statement $S1$ is
\[D^{S1} = \{(i, j) \, | \, 0 \leq i < i_{ub} \wedge 0 \leq j < j_{ub}\}\]
which represents a rectangular polyhedron in two-dimensional space bounded below by the lines $i = 0$, $j = 0$
and above by $i = i_{ub}-1$, $j = j_{ub}-1$. The corresponding instance set $I^{S1}$ is
\[I^{S1} = \{S1[i, j] \, | \,0 \leq i < i_{ub} \wedge 0 \leq j < j_{ub}\}\]

\subsection{Statement Scheduling}
In the polyhedral model \cite{poly_framework} a \textit{schedule} defines the sequential execution ordering of iterations of a
statement $S$ and is represented by an affine transformation matrix. A schedule is a mapping from statement
instances or points in an iteration space to logical \textit{time-stamps}. 
Again, the exact structure of a scheduling matrix varies between implementations and publications, 
so for clarity we describe schedules using a `mapping' notation similar to the set notation employed above.

A time-stamp is a multi-dimensional vector used to represent the logical `execution time' of a statement
instance in terms of iterators. Time-stamps are compared and ordered according to their \textit{lexicographic 
ordering} $\ll$ such that a statement instance $S1[i,j]$ is executed before $S1[i',j']$ iff 
$\theta^{S1}(i,j) \ll \theta^{S1}(i', j')$ where $i, i', j, j'$ are values of the iterators upon which
$S1$ depends.

The schedule $\theta^{S1}$ for statement $S1$ in \ref{fig:domain_snippet} is written
\[\theta^{S1} = \{S1[i, j] \rightarrow [i,j]\}\]
which demonstrates the mapping from arbitrary statement instance $S1[i,j]$ to time-stamp vector $[i, j]$.
Schedules become more interesting when dealing with loop transformations. Consider \textit{loop interchange},
a loop transformation which swaps the position of two loops in a nest and can result in better data locality by 
exploiting the layout of data in memory. This transformation could be applied to a statement 
$S = (D^S, \theta^S)$ by changing its schedule to 
\[\theta^{S1} = \{S1[i, j] \rightarrow [j,i]\}\]
which essentially `re-times' the loop nest  so that all instances of \texttt{i} occur before each subsequent 
instance of \texttt{j} and produces the transformed code in \ref{fig:interchange_snippet}. The domain of the loop
in \ref{fig:interchange_snippet} is given by $D_1^{S1}$
\[D_1^{S1} = \{ (j,i) \,|\, 0 \leq i < i_{ub} \wedge 0 \leq j \leq j_{ub}\}\]

\begin{figure}[h]
\centering
\begin{lstlisting}[language=C, style=customc]
for (int j = 0; j < j_ub; x++){
    for (int i = 0; i < i_ub; t++) {
    	S(i,j); //S1
    }
}
\end{lstlisting}
\caption{The code from \ref{fig:domain_snippet} with the \texttt{i} and \texttt{j} loops interchanged}
\label{fig:interchange_snippet}
\end{figure}

\section{Tiling}
\textit{Loop tiling}, also known as \textit{strip-mine and interchange} or \textit{loop blocking} \cite{dragon}, is an optimization that modifies 
a loop nest so that it iterates completely over `tiles' or `blocks' of its iteration space, rather than repeatedly iterating 
completely through each dimension of the iteration space until all points are visited. Tiling transformations group points of
the original iteration space into blocks. The resulting blocks have smaller iteration bounds and so the innermost loops are completed more often thereby
reducing the amount of time before nearby data accesses occur and improving data locality. The tile size can be chosen such that memory locations
required in a full loop execution fit into the cache and prevent excessive eviction. In addition to improving data locality, loop tiling can also 
create opportunities for parallelism not present in the original code.

\begin{figure}[h]
  \centering
  \subfloat[Tile size 8]{
    \includegraphics[width=0.45\textwidth]{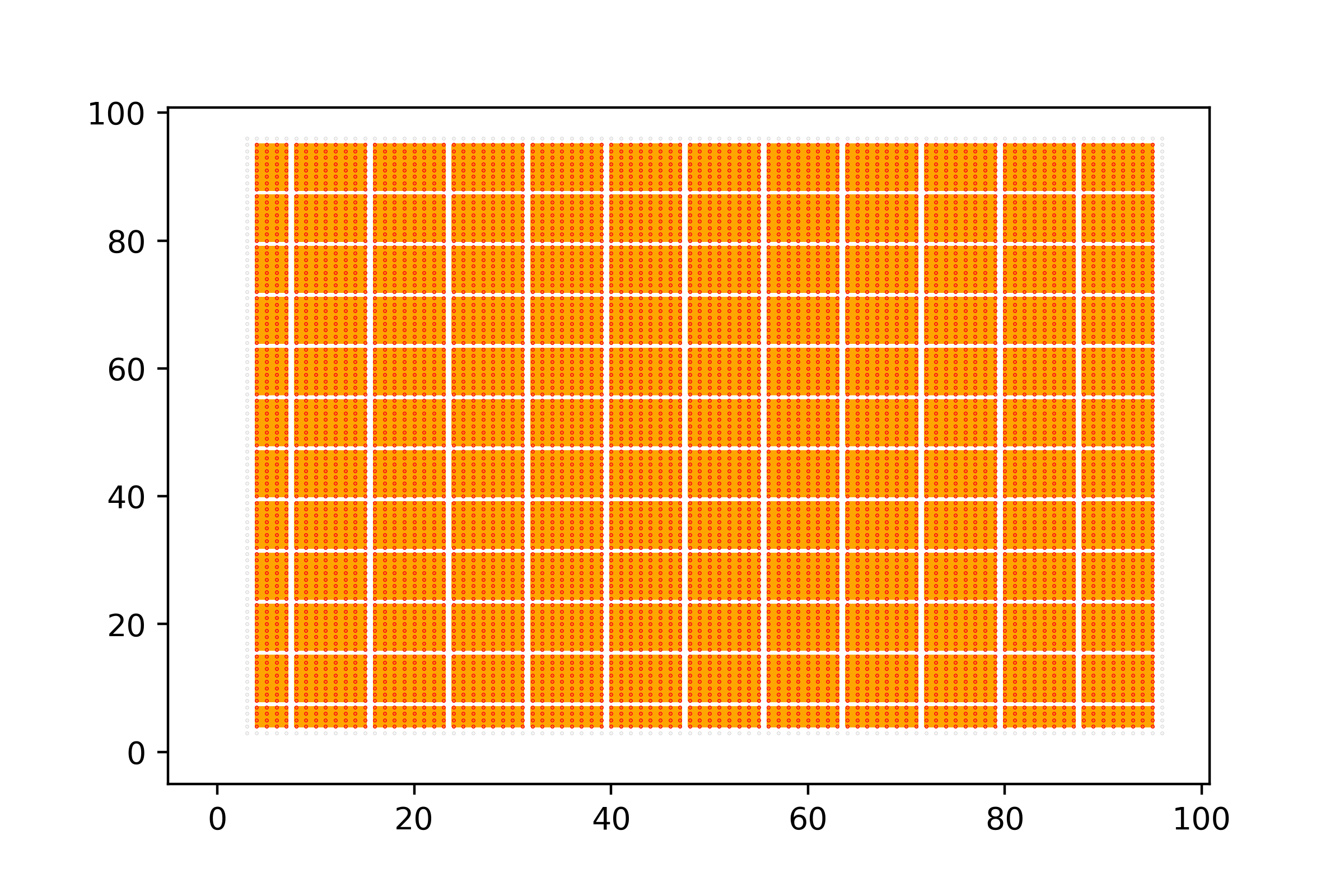}
  }
  \hfill
  \subfloat[Tile size 16]{
    \includegraphics[width=0.45\textwidth]{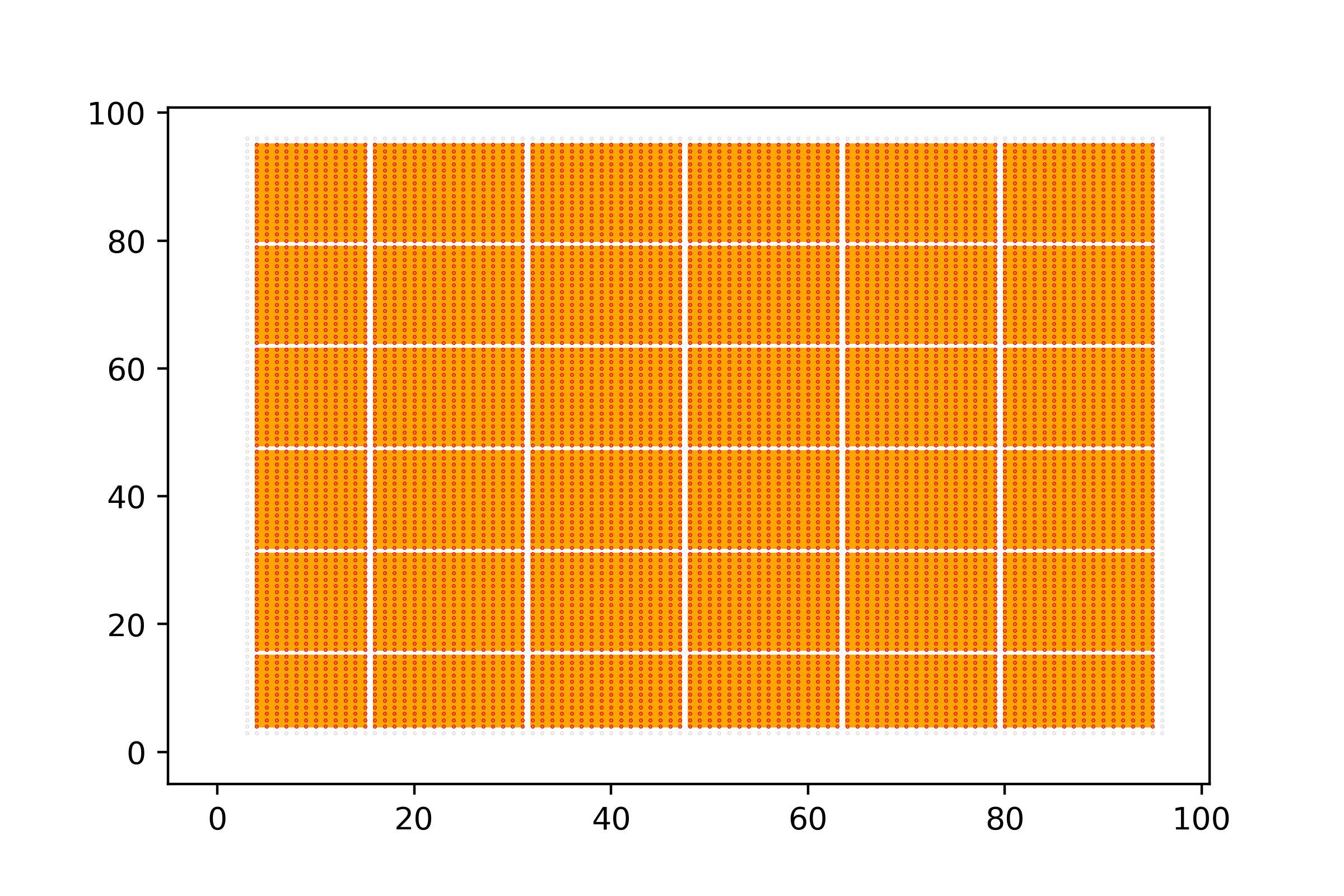}
  }
  \hfill
  \subfloat[Tile size 32]{
    \includegraphics[width=0.45\textwidth]{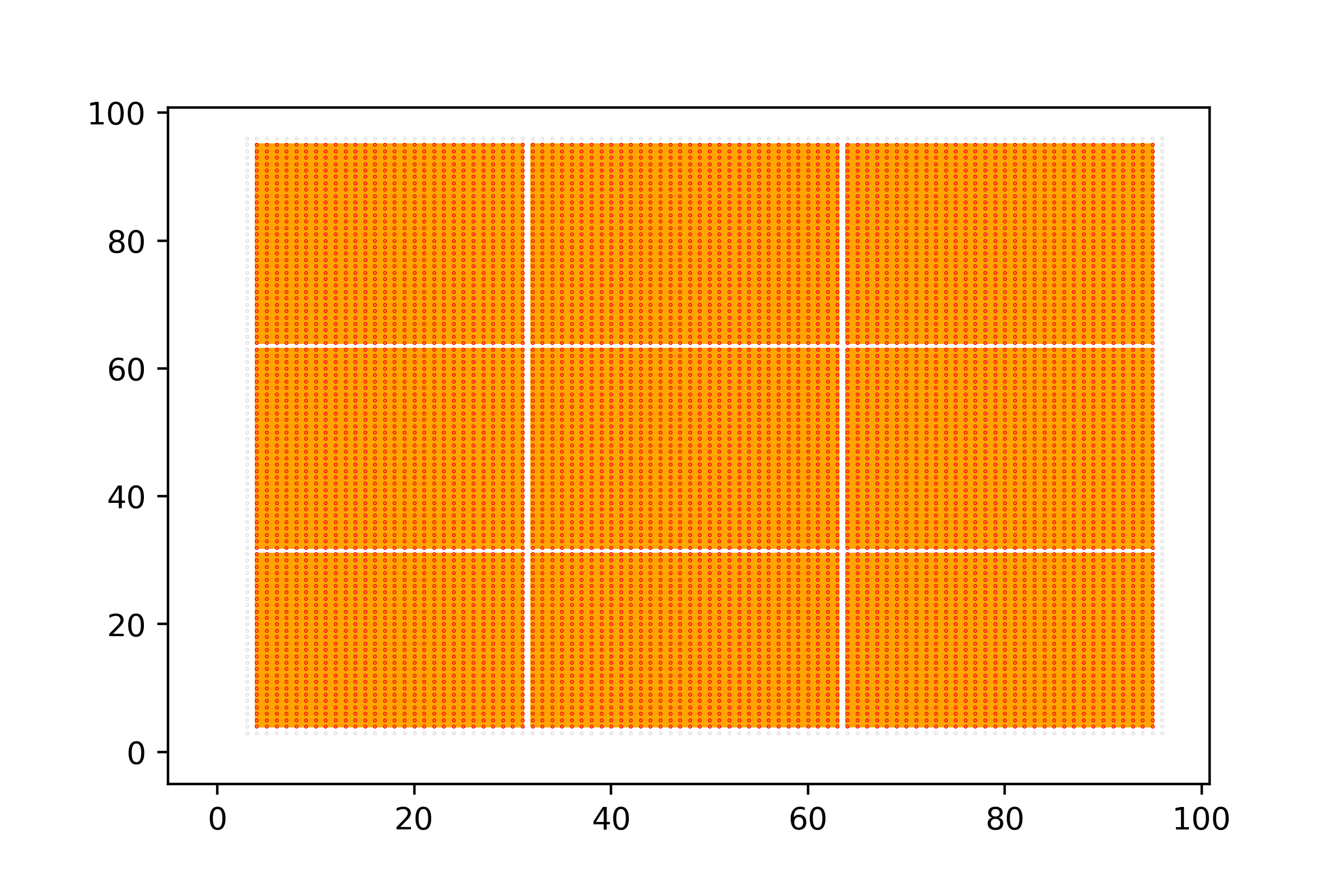}
  }
  \caption{A tiling transformation applied to the same iteration space with three different tile sizes.
  		   The grid shown is 92x92. \cite{islplot}}
  \label{fig:tile_sizes}
\end{figure}

\subsection{Implementation}
As the name \textit{strip-mine and interchange} suggests, loop tiling can be fundamentally considered 
as the composition of two basic loop transformations; \textit{strip-mining} and \textit{loop interchange}.

\subsubsection{Strip-mining}
\textit{Strip-mining} as a loop transformation takes its name from the geological mining practice
whereby large single strips of earth are removed before extracting materials from the ground below.
The basic principle in both cases centres on designating a `strip' of area and proceeding to
work on the contiguous contents of the strip before moving onto the next strip, thereby creating two sequential
processes reducing in granularity from the strip level, to the strip-contents level.
Applying this methodology to a loop requires replacing it with two loops; one outer loop to iterate 
strip-by-strip (the `strip' or `tile' loop), and one inner loop which iterates over contiguous points within the strip.
Applying this transformation to a single loop does not ultimately change the way the strip-mined loop iterates 
over its domain, in fact the transformation only serves to introduce additional control logic when compared
to the original loop. Once a loop has been strip-mined however, other loop transformations can be applied to
one or both of the resulting loops in ways which were not previously possible.

In the polyhedral representation, the domain of the loop nest in \ref{fig:stripmine_snippet} has been transformed from
\[D^{S1} = \{ (i,j) \,|\, 0 \leq i < i_{ub} \wedge 0 \leq j \leq j_{ub}\}\]
to
\[D_1^{S1} = \{ (i,jj,j) \,|\, 0 \leq i < i_{ub} 
         \,\wedge\, jj \leq j < j_{ub} 
         \,\wedge\, jj \leq j < \texttt{T}*jj + \texttt{T}
         \,\wedge\, 0 \leq jj < \floor*{\frac{j_{ub}}{\texttt{T}}}\} \]
         
\begin{figure}[h]
\centering
\begin{lstlisting}[language=C, style=customc]
for (int i = 0; i < i_ub; x++){
    for (int jj = 0; j < floord(j_ub, T); t++) {
        for (int j = jj; j < min(j_ub, T*jj + T); t++) {
            S(i,j); //S1
        }
    }
}
\end{lstlisting}
\caption{The code from \ref{fig:domain_snippet} but the \texttt{j} loop has been strip-mined.
The new \texttt{jj} loop is the `strip' or `tile' loop and the tile size is \texttt{T}.}
\label{fig:stripmine_snippet}
\end{figure}

\begin{figure}[h]
\centering
\includegraphics{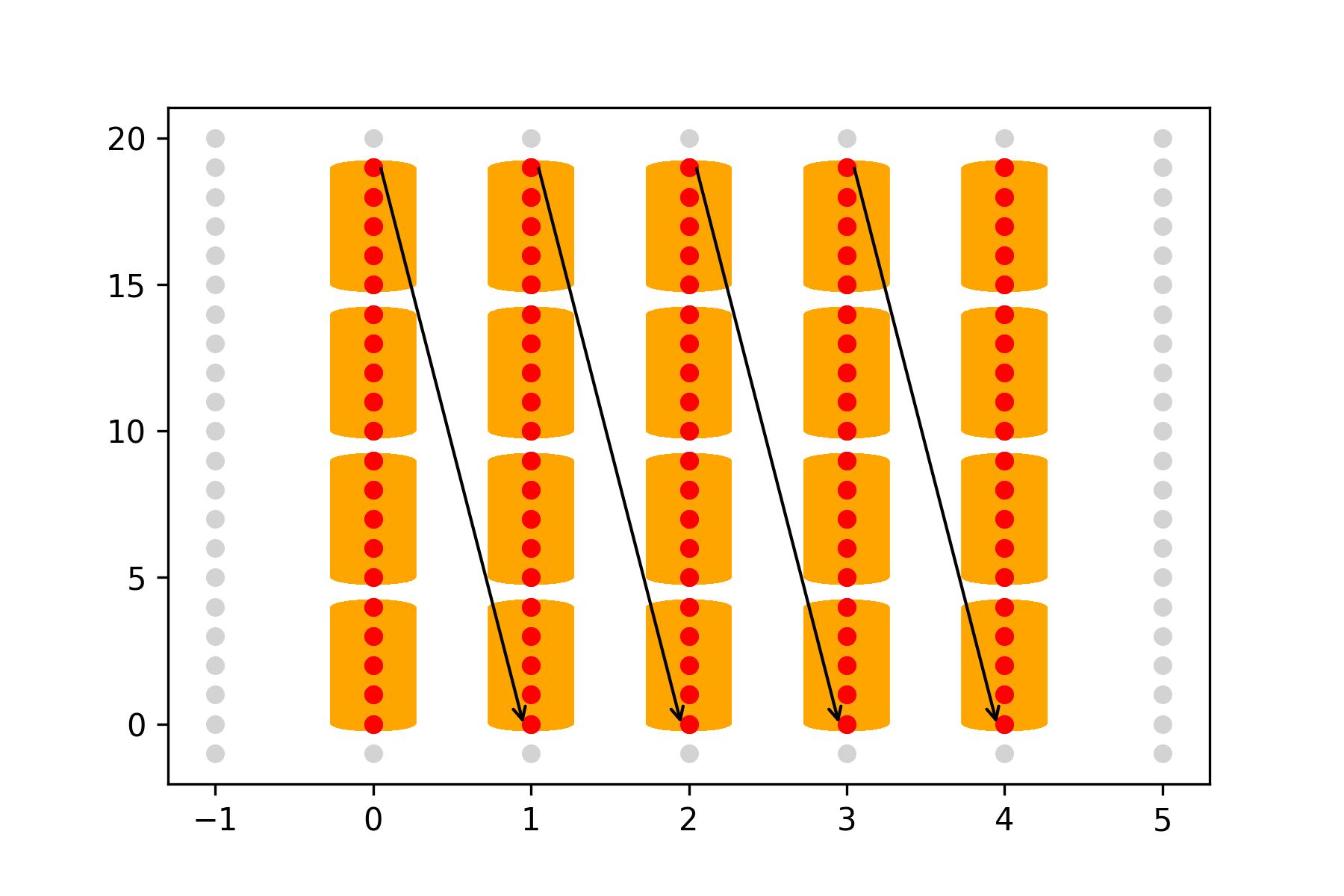}
\caption{A visualization of the iteration space for \ref{fig:stripmine_snippet}. The 
horizontal and vertical axes represent iterations in the \texttt{i} and \texttt{j} dimensions respectively.
Tiles are highlighted in orange. \texttt{i\_ub}, \texttt{j\_ub} and \texttt{T} are 5, 20 and 5
respectively. Since the \texttt{jj} loop remains inside the 
\texttt{i} loop, a full column of the iteration space (containing four tiles in this case)
must be executed before moving to the next column (by incrementing \texttt{i}), 
making the order of iteration identical to the loop before strip-mining.
The arrows indicate the order and direction of iteration. \cite{islplot}}
\label{fig:stripmine_space}
\end{figure}

\clearpage
\subsubsection{Loop Interchange}
\textit{Loop interchange} refers to the practice of `swapping' or interchanging the position of two loops in a loop nest
so that relative to each other, the outer loop becomes the inner and vice versa. The effects of this
can be seen in the transformation from \ref{fig:domain_snippet} to \ref{fig:interchange_snippet}.
Loop interchange is one of the most basic loop transformations possible, but its impact on performance and 
logical behaviour can be significant as it changes the order iteration points are accessed significantly.
Loop interchange is often used to ensure that memory accesses within a loop nest proceed according to the layout of data
in memory e.g to change the accesses of a two-dimensional loop nest from row-by-row to column-by-column.

\begin{figure}[h]
  \centering
  \subfloat[Before loop interchange]{
    \includegraphics[width=0.45\textwidth]{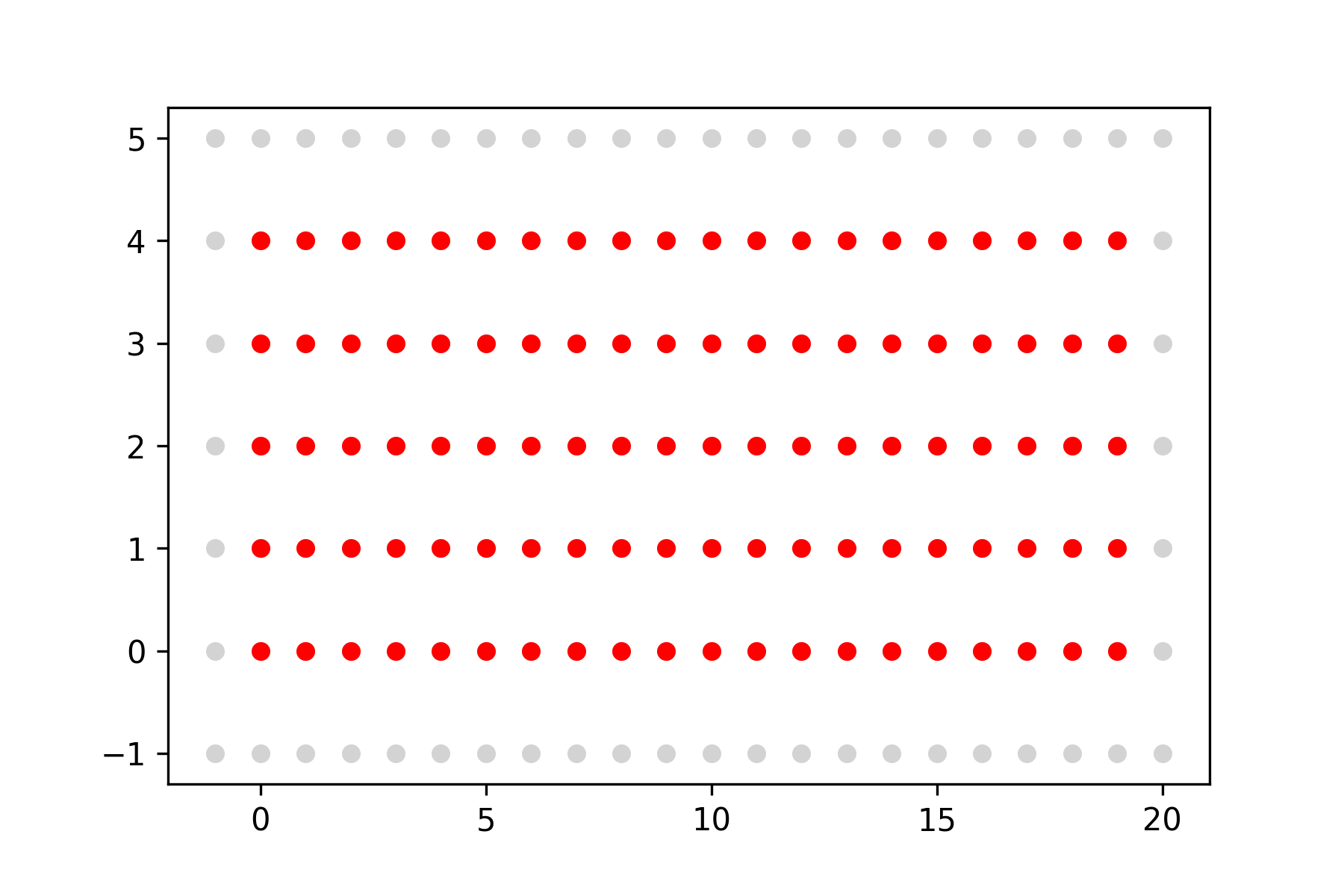}
  }
  \hfill
  \subfloat[After loop interchange]{
    \includegraphics[width=0.45\textwidth]{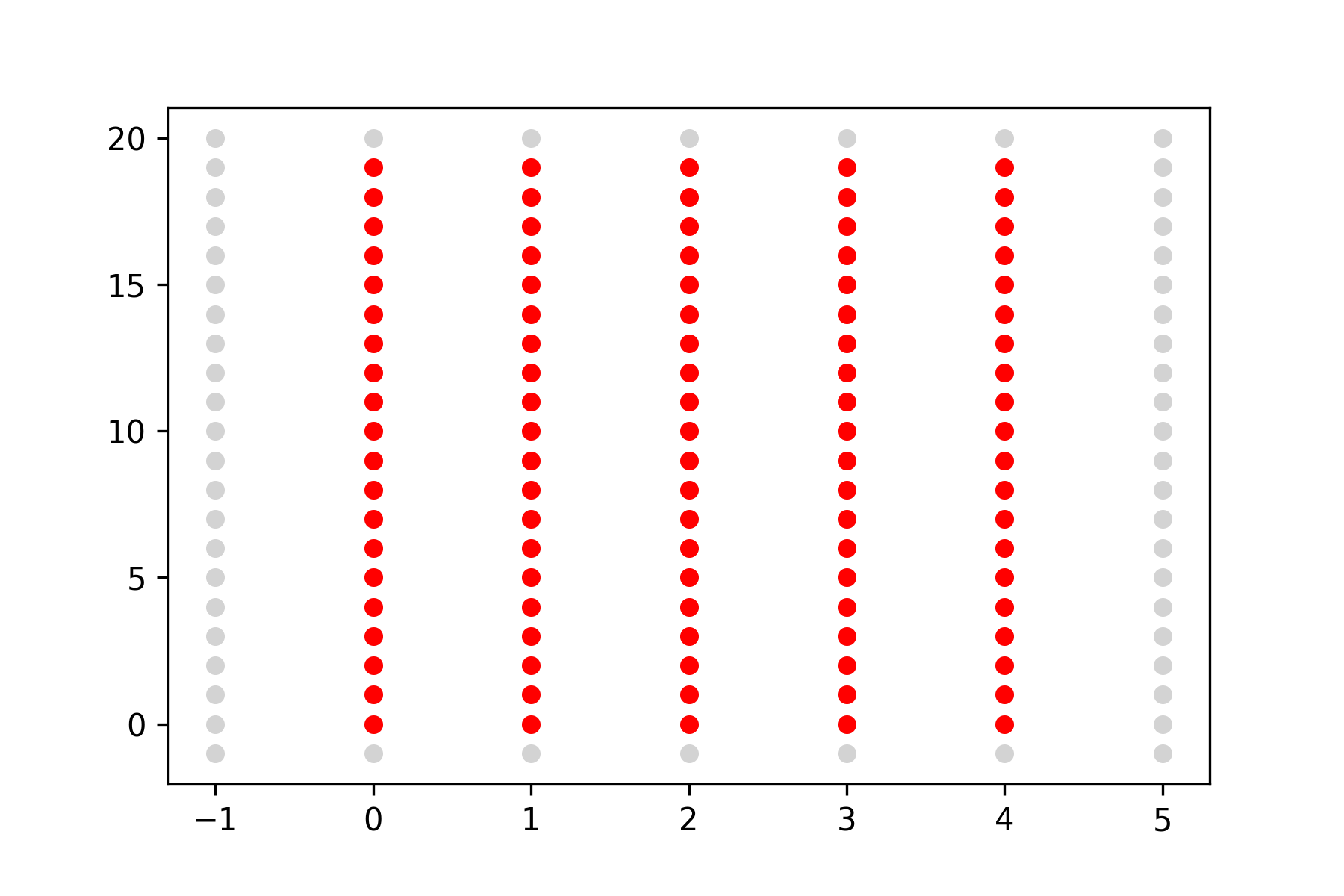}
  }
  \caption{The iteration spaces for \ref{fig:domain_snippet} and \ref{fig:interchange_snippet}
  after setting \texttt{i\_ub} and \texttt{j\_ub} to 20 and 5 respectively.
  Iteration proceeds sequentially from one column to the next. The red dots represent
  iteration points or statement instances. \cite{islplot}}
\end{figure}

\subsubsection{Combining strip-mining and loop interchange}
Combining \textit{strip-mining} and \textit{loop interchange} on one or more loops in a loop nest results in
a tiled iteration space. This is achieved by interchanging the `tile' loop so that it is outside of one or more
enclosing loops. Since the tile loop is now an outer loop, and the inner tiled loop bounds are defined in terms of the
tile loop's iterator, the new loop nest proceeds one tile at a time (a single iteration of the tile loop) and executes
all iterations within that tile's bounds before moving on to the next tile. Consider the code in \ref{fig:stripmine_snippet},
by applying a loop interchange transformation which exchanges the order of the \texttt{i} and \texttt{jj} loops we obtain the code
in \ref{fig:tiled_snippet}. We also set the values of \texttt{i\_ub}, \texttt{j\_ub} and \texttt{T} to 10, 20 and 5 respectively
to demonstrate the tiled iteration space in \ref{fig:tiled_space}. This interchange transformation is represented by the
polyhedral schedule $\theta^{S1}$
\[\theta^{S1} = {S1[i,jj,j] \rightarrow S1[jj,i,j]}\]

\begin{figure}[h]
\centering
\begin{lstlisting}[language=C, style=customc]
for (int jj = 0; j < floord(20, T); t++) {
    for (int i = 0; i < 10; x++){
        for (int j = jj; j < min(20, 5*jj + 5); t++) {
            S(i,j); //S1
        }
    }
}
\end{lstlisting}
\caption{Tiled code produced by applying loop interchange to the \texttt{i} and \texttt{jj} loops
in \ref{fig:stripmine_snippet}, and setting the values of \texttt{M}, \texttt{N} and \texttt{T} to 10, 20 and 5 respectively.}
\label{fig:tiled_snippet}
\end{figure}

\begin{figure}[h]
\centering
\includegraphics{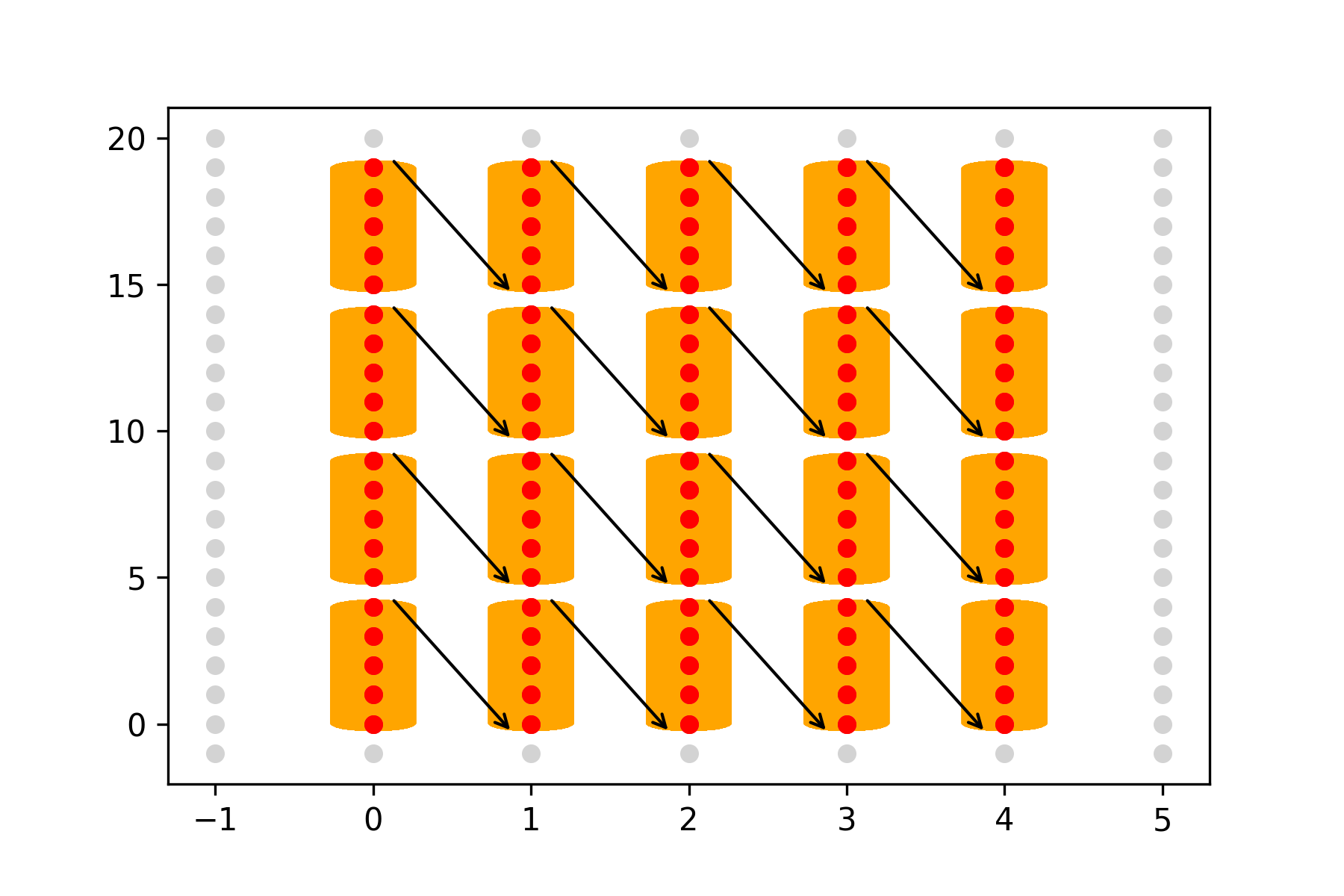}
\caption{A visualization of the iteration space for \ref{fig:tiled_snippet}.
Tiles are highlighted in orange. As tiling was only performed in one dimension,
the resulting tiles are essentially one-dimensional strips which are each executed 
sequentially in a row before moving to the next row (equivalent to incrementing the \texttt{jj} loop) as indicated by the arrows from 
tile to tile. \cite{islplot}}
\label{fig:tiled_space}
\end{figure}

\section{Legality of Tiling}
The dependence distance vectors of a loop nest enforce a partial order on the execution of its iterations \cite{tiling_legality}.
Tiling transformations may re-order iterations but in order to be legal transformations which
preserve the logical behaviour of the underlying code, dependencies must be respected. 

In general, a loop transformation is legal iff the data dependence vectors of the resulting code are lexicographically
positive (the first non-zero element of the vector is positive) \cite{tiling_legality}. Otherwise it would be possible to have a loop nest 
where in one dimension an earlier iteration depends upon a later one (i.e there is a negative dependence 
distance in that dimension) while any outer loops remain at the same iteration in their own dimensions 
(i.e dependence distances of zero in all dimensions before the negative dependence carrying one). 
This would create a dependence on future iterations which have not yet been completed.

\begin{figure}[h]
\centering
\includegraphics[scale=0.85]{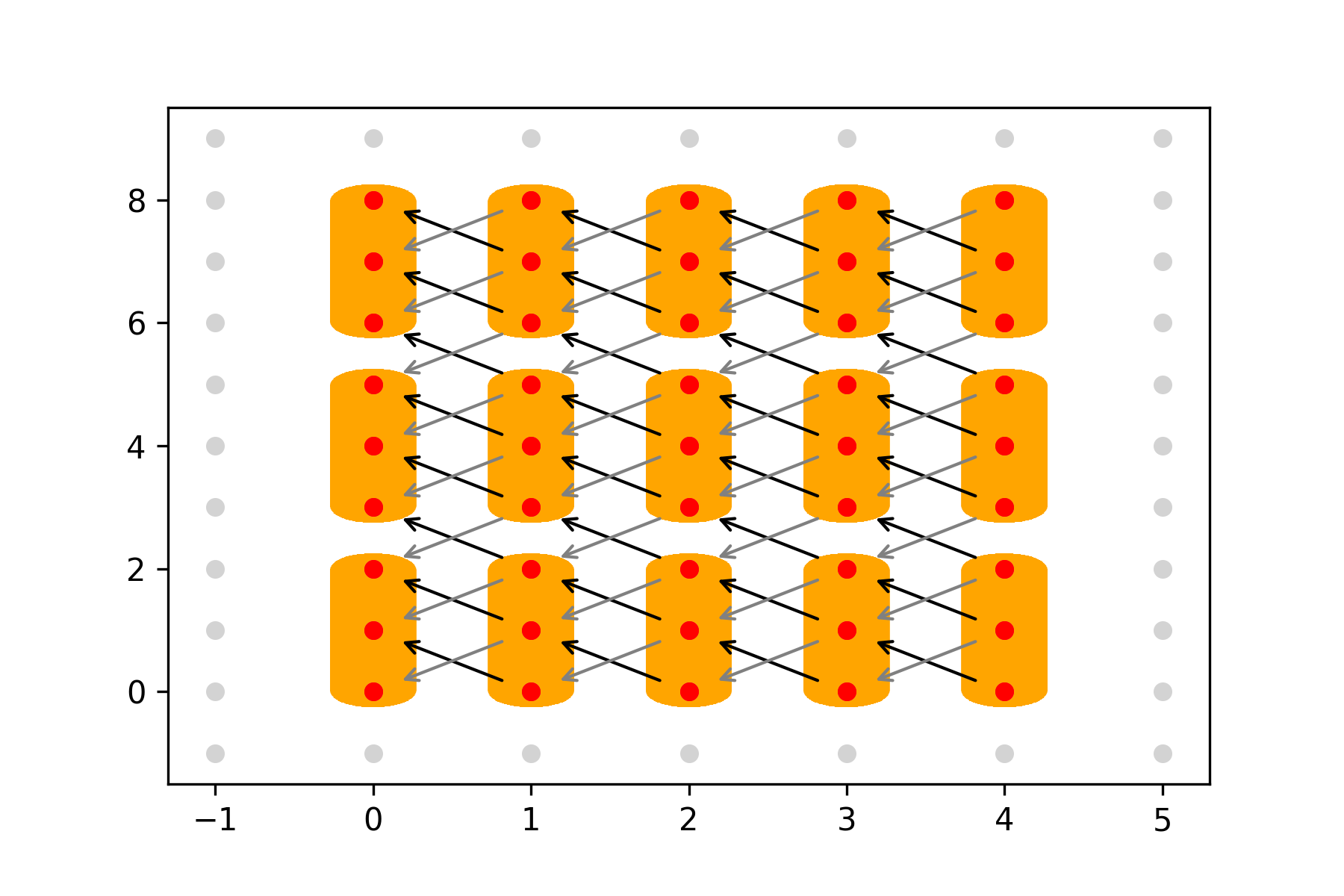}
\caption{The iteration space for \ref{fig:noskew_snippet} after applying an illegal tiling transformation
with a tile size of 3. The grey and black arrows represent dependences. \cite{islplot}}
\label{fig:illegal_space}
\end{figure}

Strip-mining is not an iteration re-ordering transformation so dependence distance vectors do not change
after its application. Loop interchange however, exchanges rows of the original dependence distance vectors
and so is not always a legal transformation. Since loop interchange is a fundamental component of tiling transformations,
tiling transformations can also be illegal by creating tiles which depend on subsequent tiles that have yet to be executed.
Consider the code in \ref{fig:noskew_snippet}, after applying an illegal rectangular tiling transformation we obtain the iteration
space in \ref{fig:illegal_space} where all tiles in a row are executed sequentially before moving to the next row. In this
case, the black arrow represents the data dependence whose dependence distance is negative. We can see by looking
at any row of tiles that there is a black dependence arrow pointing from the last point in each tile to the first point
in a subsequent tile located one row above, which is an illegal dependence on an iteration not yet executed.

The basic strip-mine and interchange operations applied to produce \ref{fig:illegal_space}
were not inherently incorrect transformations, rather the dependence structure to which
they were applied required additional transformations beforehand so that the resulting 
data dependencies were legal. Transformations like loop skewing can be used to remove
negative dependence distances and `legalize' further transformations.

\section{Skewed Tiling}
In order to tile stencil codes with negative data dependence distances in
one or more of the tiled dimensions, the iteration space must first 
be manipulated such that iteration points in one tile do not depend on iteration points 
in a subsequent tile. This can be achieved through loop skewing, a transformation
by which each iteration in the skewed dimension is `skewed' or `offset' by the product 
of the enclosing loop's iterator and a chosen skewing factor, effectively
offsetting dependence distances in the skewed dimension and potentially making them positive.

\subsection{Loop Skewing}
Skewing an iteration space is analogous to skewing an image or polyhedron on a graph, indeed 
iteration spaces are convex polyhedra (or unions of polyhedra) in the polyhedral model.  A loop skewing transformation
changes the bounds of an inner loop to depend on the outer loop with respect to which it is being skewed so that
with each iteration of the outer loop, the skewed loop is `shifted' further along its dimension. Often,
skewing transformations are used to expose opportunities for parallelism, or enable further loop
transformations that would not be possible without skewing.
The transformation and its benefits can be more easily understood with the code snippets and corresponding 
iteration spaces shown below.

\begin{figure}[h]
\centering
\begin{lstlisting}[language=C, style=customc]
for (int t = 0; t < M; t++) {
	for (int x = 0; x < N; x++){
    	a[t][x] = a[t-1][x-1] + a[t-1][x+1];
    }
}
\end{lstlisting}

\caption{A simple two-dimensional stencil loop with dependencies carried along the \texttt{t} loop.}
\label{fig:noskew_snippet}
\end{figure}

The loop nest in \ref{fig:noskew_snippet} features two `true dependencies' carried along the \texttt{t} loop
which are induced by two flow of data read from \texttt{a[t-1][x-1]} and \texttt{a[t-1][x+1]} into the assignment 
\texttt{a[t][x]}. The distance vectors for these two dependencies are, respectively, 
$\vec{d}_1 = (1, 1)$ and $\vec{d}_2 = (1, -1)$.

\begin{figure}[h]
\centering
\begin{lstlisting}[language=C, style=customc]
for (int t = 0; t < M; t++) {
	for (int x = 2*t; x < N + 2*t; x++){
    	a[t][x-2*t] = a[t-1][x-1-2*t] + a[t-1][x+1-2*t];
    }
}
\end{lstlisting}
\caption{The stencil from \ref{fig:noskew_snippet} but the \texttt{x} loop has been skewed by a factor of 2 relative to
the \texttt{t} loop. Note that the bounds of the \texttt{x} loop and expressions in the body containing
\texttt{x} have been modified to include the `skewing term' \texttt{2*time}. The transformed indices
in the loop body correspond to a translation from the skewed iteration space of the loop to the
original array index space to preserve logical correctness.}
\label{fig:skew_snippet}
\end{figure}

\begin{figure}[h]
  \centering
  \subfloat[Before loop skewing]{
    \includegraphics[width=0.45\textwidth]{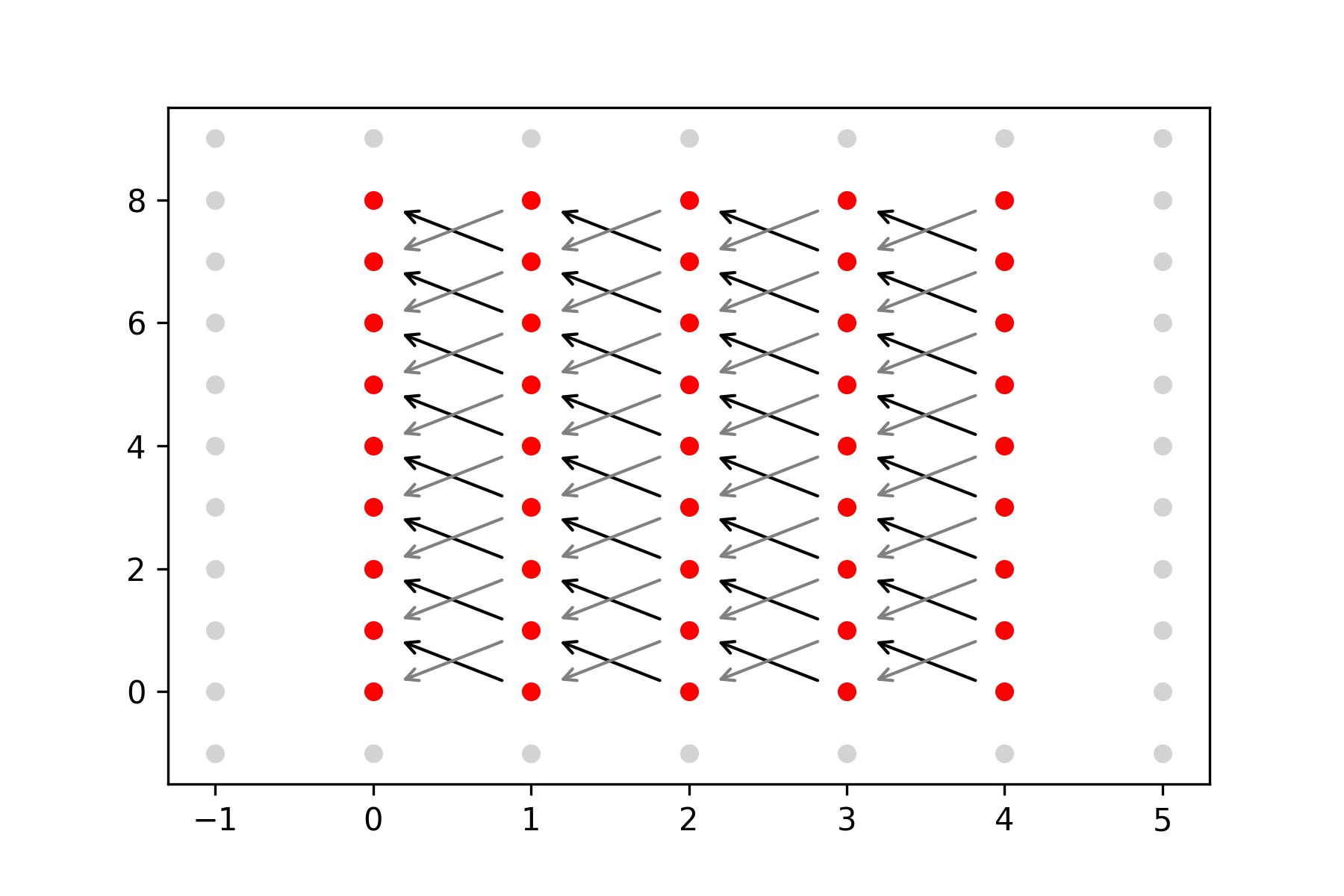}\label{fig:noskew_space}
  }
  \hfill
  \subfloat[After loop skewing]{
    \includegraphics[width=0.45\textwidth]{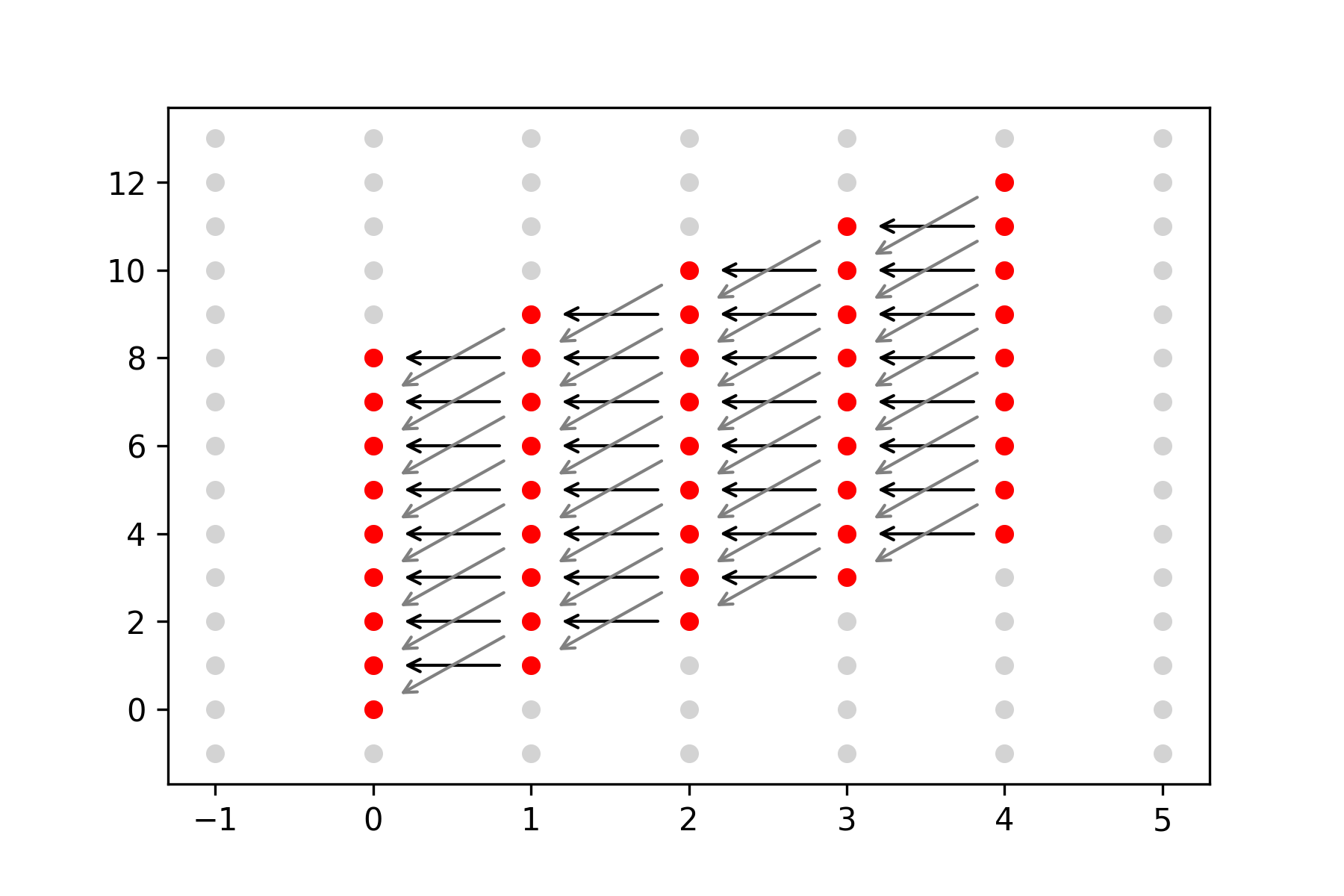}\label{fig:skew_space} 
  }
  \caption{The iteration spaces for \ref{fig:noskew_snippet} and \ref{fig:skew_snippet}
  after setting \texttt{M} and \texttt{N} to 9 and 5 respectively.
  Iteration proceeds sequentially from one column to the next. The grey and black arrows
  indicate data dependencies present in the loop body. Arrows point outward from one iteration point
  to all other points which calculate values it depends on. In this case the grey and black arrows 
  correspond to the dependencies introduced by the \texttt{a[t-1][x-1]} and \texttt{a[t-1][x+1]}
  terms respectively. \cite{islplot}}
\end{figure}

In \ref{fig:skew_snippet}, the skewing transformation enables a previously impossible
rectangular tiling transformation by skewing the dependencies introduced by the \texttt{a[t-1][x+1]}  
term so they point `horizontally' between \texttt{t} iterations as in \ref{fig:skew_space}, 
rather than diagonally as in \ref{fig:noskew_space}. This is equivalent to increasing the dependence distances
in the \texttt{x} dimension by 1 so that they are all non-negative, producing dependence distance vectors
$\vec{d'}_1 = (1, 2)$ and  $\vec{d'}_2 = (1, 0)$.
Once skewed in this way, an iteration point
only depends on other iteration points which are no further along the \texttt{x} dimension (i.e
no dependence arrows are pointing diagonally upward), permitting the tiling transformations shown
in \ref{fig:tiled_skew_snippet}, \ref{fig:tiled_skew_space}.

Executing \ref{fig:skew_snippet} as it is presented above would not provide any performance gain over 
\ref{fig:noskew_snippet}
(indeed the complex loop bounds introduced in the skewing transformation may make performance worse
than its non-tiled counterpart), however by enabling further optimizing transformations 
\ref{fig:skew_snippet} is a first step towards code with better data locality.

\subsubsection{Formal description}
Skewing alone does not affect the way in which a loop iterates over its own domain, instead the position 
of the loop's iterations is changed relative to other loops around or within it. This can be seen
in how the shape of a skewed loop's iteration domain (the columns of \ref{fig:noskew_space}, 
\ref{fig:skew_space}) is constant, but it is shifted relative to the other dimensions in the loop nest
(the horizontal \texttt{t} axis in \ref{fig:noskew_space}, \ref{fig:skew_space}). 
Using the polyhedral model, the skewing transformation for a statement $S1$ can be represented by the 
schedule in \ref{fig:skew_schedule}

\begin{figure}[h]
\centering
\[\theta^{S1} = \{S1[t, x] \rightarrow S[t, x + kt]\}\]
\[D^{S1} = \{[t, x] \,|\, t_{lb} \leq t < t_{ub} \,\wedge\, x_{lb} \leq x < x_{ub}\}\]
\[D_1^{S1} = \{[t, x] \,|\, t_{lb} \leq t < t_{ub} \,\wedge\, x_{lb} + kt \leq x < x_{ub} + kt\}\]
\caption{A generic scheduling transformation for skewing a loop $x$ with respect to a loop $t$ by a 
skewing factor $k$. $D^{S1}$ and $D_1^{S1}$ are the domains before and after skewing respectively.}
\label{fig:skew_schedule}
\end{figure}

\subsection{Combining with Tiling}
Skewing a loop can unlock opportunities for tiling that were not present in the original loop
due to its dependence structure. Skewing a loop and then applying a tiling transformation is called
\textit{skewed tiling}. The order of these two transformations is very important. Tiling a loop nest and then skewing
it with the schedule in \ref{fig:skew_schedule} results in tiled loops whose bounds are then skewed only to be `unskewed' in their
bodies so that indexing refers to the original iteration space. The transformation produces tile shapes
which are skewed relative to their original shape but maintain their original dimensions making the skewing transformation
redundant. 

\begin{figure}[h]
\centering
\includegraphics{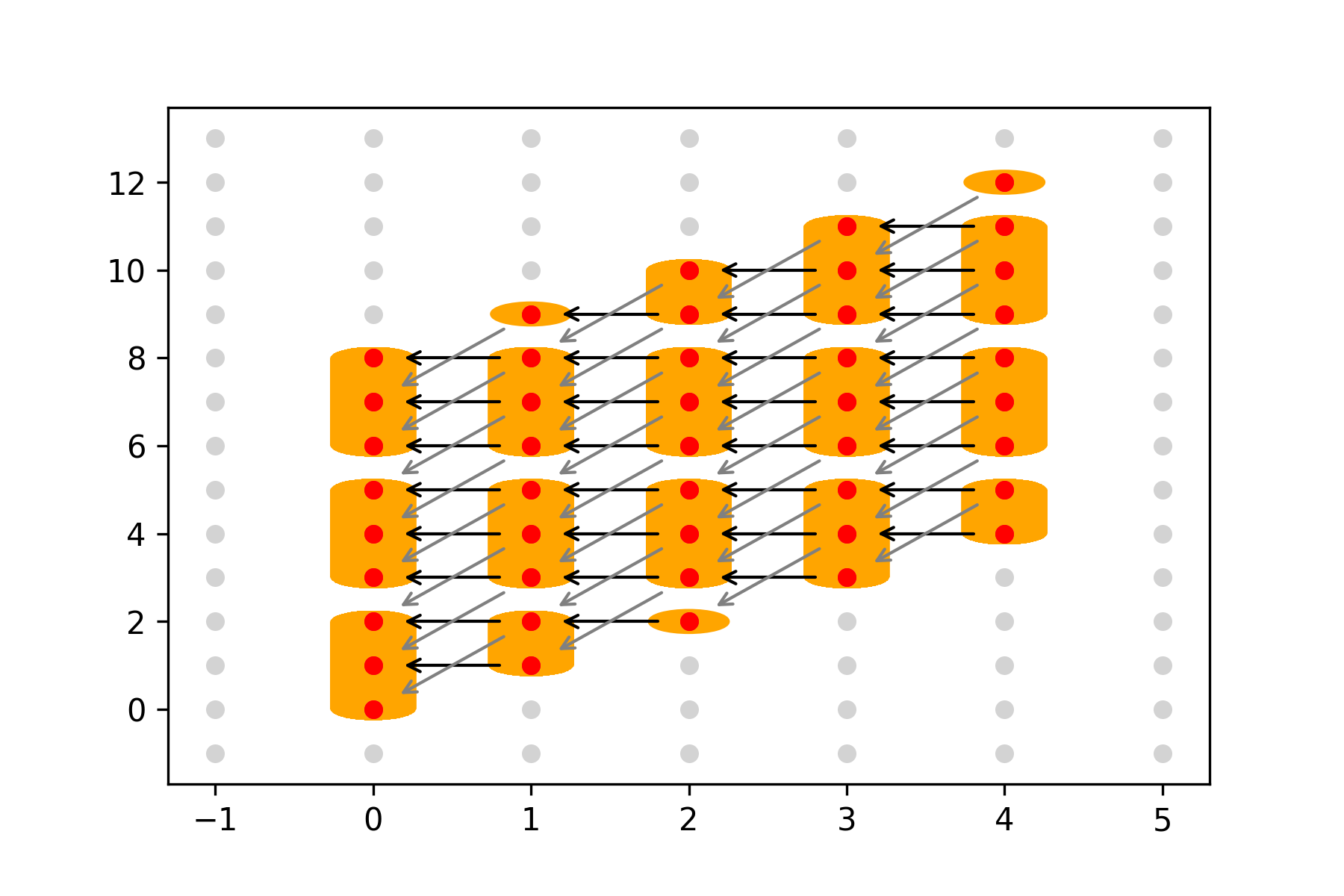}
\caption{The iteration space from \ref{fig:skew_space} after applying a transformation
to tile the \texttt{x} loop with a tile size of 3. Note that because of the interchange
transformation which places the \texttt{xx} tile loop outside the \texttt{t} and \texttt{x} loops,
iteration proceeds sequentially across all tiles in a row before moving to the next row of tiles
(as described in \ref{fig:tiled_space}). \cite{islplot}}
\label{fig:tiled_skew_space}
\end{figure}

Applying a skewing transformation before tiling a loop nest means the tiling transformation operates
on skewed loop domains. In this case tiles maintain the shape given by the tiling configuration and are not skewed,
however because the underlying iteration space is skewed, some tiles may not be whole tiles with regular edges. 
A basic rectangular tiling transformation of the code in \ref{fig:skew_snippet} and the iteration space in \ref{fig:skew_space}
gives \ref{fig:tiled_skew_snippet} and \ref{fig:tiled_skew_space}.

\begin{figure}[h]
\centering
\begin{lstlisting}[language=C, style=customc]
for(int xx = 0; xx < floord(N, T); x++){
for (int t = 0; t < M; t++) {
	for (int x = T*xx + 2*t; x < min(N+ 2*t, T*xx + T); x++){
    	a[t][x-2*t] = a[t-1][x-1-2*t] + a[t-1][x+1-2*t];
    }
}
\end{lstlisting}
\caption{The stencil from \ref{fig:skew_snippet} with a \textit{skewed tiling} transformation applied.}
\label{fig:tiled_skew_snippet}
\end{figure}

The tiling transformation presented in \ref{fig:tiled_skew_snippet}, \ref{fig:tiled_skew_space} is
valid because no tile depends on values computed in any subsequent tile. This can be seen in
\ref{fig:tiled_skew_space} by how no tile has dependence arrows which point to any subsequent tile
along the tiled dimension and so tiles can be executed across multiple \texttt{t} iterations while
respecting data dependencies 

\section{Summary}
We have introduced general-purpose loop tiling as a transformation in the polyhedral model and shown
that it is a composition of two loop transformations; strip-mining and loop interchange. Adding 
loop skewing to this combination can result in skewed tiling, a more advanced transformation which can be
applied to loops which would not ordinarily be suitable for tiling. In the next chapter we take the fundamental
tiling transformations discussed above, apply them to the space-time loops seen in Devito, and introduce a
further transformation; \textit{time-tiling}.

\chapter{Time-tiling for Devito stencils}
\label{ch:timetiling}
In this chapter we introduce the primary optimization researched in this project, \textit{time-tiling},
and relate it to the spatial tiling techniques currently applied by Devito.
This tiling transformation applies to space-time loops and creates iteration space tiles which contain multiple
iterations in the outermost time dimension resulting in loop nests with improved data reuse
in all dimensions. The following discussion is applicable to stencil loops with uniform
dependencies which are fixed throughout program execution, and known at compile time.

Tiling through the dependence-carrying time dimension requires additional iteration space 
transformations which depend on the dependence 
structure of the stencil and is not currently implemented in Devito. 
Stencil codes in Devito carry dependencies along the time dimension
as they update points in the spatial dimensions based on weighted contributions from nearby spatial points at 
previous time steps. In order to tile the time dimension and respect these dependencies, a skewed tiling
transformation must be applied to the stencil loop.

\section{Space Tiling in Devito}
In Devito's spatial tiling transformations, all but the innermost
space loops are tiled according to the same tiling principles described in the previous chapter but without 
knowledge of the loop bounds. General-case parameterized loops are generated which explicitly
handle edge cases for situations where the iteration space cannot be partitioned entirely into whole tiles.
Since tiling is only applied to space loops and Devito stencil codes only have dependencies carried along
the time dimension, no additional transformation of the iteration space is required for tiling to be legal.
This manner of tiling takes advantage of data locality in the space dimensions but does not benefit from
reuse of data in subsequent time iterations.

\begin{figure}[h]
\centering
\begin{lstlisting}[language=C, style=customc]
for (int time=1; time < time_size-1; time++){
  for (int x_b=4; x_b < x_size-(x_size-8)%(x_b_size)-4; x_b += x_b_size){
    for (int y_b=4; y_b < y_size-(y_size-8)%(y_b_size)-4; y_b += y_b_size){
      for (int x=x_b; x < x_b + x_b_size; x++){
        for (int y=y_b; y < y_b + y_b_size; y++){
          for (int z=4; z < z_size-4; z++){
            //Stencil code
          }
        }
      }
    }
  }
  for (int x=x_size-(x_size-8)%(x_b_size)-4; x < x_size-4; x++){
    for (int y=4; y < y_size-(y_size-8)%(y_b_size)-4; y++){
      for (int z=4; z < z_size-4; z++){
        //Stencil code
      }
    }
  }
  for (int x=4; x < x_size-(x_size-8)%(x_b_size)-4; x++){
    for (int y=y_size-(y_size-8)%(y_b_size)-4; y < y_size-4; y++){
      for (int z=4; z < z_size-4; z++){
        //Stencil code
      }
    }
  }
  for (int x=x_size-(x_size-8)%(x_b_size)-4; x < x_size-4; x++){
    for (int y=y_size-(y_size-8)%(y_b_size)-4; y < y_size-4; y++){
      for (int z=4; z < z_size-4; z++){                                                                                                                          
        //Stencil code
      }
    }
  }
}
\end{lstlisting}
\caption{Devito cache-blocked code with tiled \texttt{x} and \texttt{y} dimensions.
Formatted for clarity.}
\label{fig:cache_block_snippet}
\end{figure}

\section{Skewing Dependencies}
The acoustic wave equation stencil codes which were the primary targets for our transformations
feature loop-carried dependencies which prevent regular tiling transformations from being applied.
Depending on the \textit{space order} (a parameter which affects how many nearby spatial points are considered
in the stencil code) of the simulation being run, the AWE stencil updates an array with data from previous time steps
at both previous/subsequent spatial points. Since the dependence is carried along the time dimension,
there are no loop-carried dependencies in the spatial loops and so they can be readily tiled.
Applying rectangular tiling to the whole loop nest as-is would result in iteration points attempting to 
read from subsequent spatial points at previous time steps which are located in tiles that have not yet been executed
as demonstrated (via a more simple example) in \ref{fig:illegal_space}.

\begin{figure}[h]
\begin{lstlisting}[language=C, style=customc]
  u[time][x][y][z] = ... 
 -8.25142857142857e-5F*(u[time-1][x][y][z - 4] //d1
                       +u[time-1][x][y][z + 4] //d2
                       +u[time-1][x][y - 4][z] //d3
                       +u[time-1][x][y + 4][z] //d4
                       +u[time-1][x - 4][y][z] //d5
                       +u[time-1][x + 4][y][z] //d6
                       ) 
                     ... ;
\end{lstlisting}
\caption{The left hand side and an isolated term from the right hand side of the acoustic wave equation stencil calculation.}
\label{fig:stencil_dependencies}
\end{figure}

In \ref{fig:stencil_dependencies} the dependencies which prevent tiling are `true dependencies' introduced 
by terms which index into the spatial dimensions with a positive offset (i.e \texttt{u[time-1][x][y + 4][z]}).
For the array accesses terms in \ref{fig:stencil_dependencies} the dependence distance vectors are
\[\vec{d}_1 = (1, 0, 0, 4) \quad \vec{d}_2 = (1, 0, 0,-4)\]
\[\vec{d}_3 = (1, 0, 4, 0) \quad \vec{d}_4 = (1, 0,-4, 0)\]
\[\vec{d}_5 = (1, 4, 0, 0) \quad \vec{d}_6 = (1,-4, 0, 0)\]
with $\vec{d}_2$ and $\vec{d}_4$ introducing the negative dependence distances which prevent tiling.
To solve this, each loop corresponding to a dimension along which a negative dependence exists must be 
skewed relative to the outermost loop by a factor equal to the negated value of greatest negative dependence in that dimension.
This effectively `straightens' all dependencies in each dimension
so that all dependence distance vectors have a distance greater than or equal to 0 in the skewed dimension,
producing an change in iteration space similar to \ref{fig:skew_space}. In the case of the AWE stencils
evaluated in this research, the \texttt{z} dimension is not tiled so only the \texttt{x} and \texttt{y} loops
require skewing.

\section{Time-Tiling}
Time-tiling brings the temporal and spatial locality improvements offered by tiling transformations
to the time dimension of a space-time loop. Time-tiling can significantly reduce the time between data reuses as the stencil 
moves through time and shares many data points with previous time steps. \ref{fig:time_tiled} and
\ref{fig:dep_tiled} demonstrate the effect of these transformations on an iteration space, and visualize 
the locality improvements it brings about.

Unlike skewed tiling, time-tiling is not a general-purpose transformation with a formal definition
as its implementation code varies depending on the dependencies in the loop structure being optimized. 
Fundamentally, \textit{time-tiling} refers to a composition of loop transformations which when applied to a 
space-time loop nest featuring an outer time loop and multiple inner space loops, produces
code which iterates through the spatial dimensions in a tiled fashion, and contains multiple time steps 
in a tile. This does not necessarily require strip-mining of the time loop, but at a minimum the time loop
must be interchanged so that it is within the spatial tile loops, creating tiles which extend fully through the
time dimension (Tilings of this kind are evaluated in \ref{fig:256_save_results}). More advanced time-tiling
transformations may tile the time loop as well, so that a spatial tile executes a fixed number of time steps
that is less than the total number of time steps.

The challenge with time-tiling is that
applications which benefit from it typically have complex dependence structures and work must be done
to prepare the iteration space so that tiling the time dimension respects those dependencies. In the case
of the acoustic wave equation stencil used in our evaluation of time-tiling, this preparation amounts
to skewing the iteration space so that all dependence distance vectors are lexicographically positive.

\section{Comparison with space-tiling}

\begin{figure}[h]
  \centering
  \subfloat[Normal Devito stencil with no tiling]{
    \includegraphics[width=0.6\textwidth]{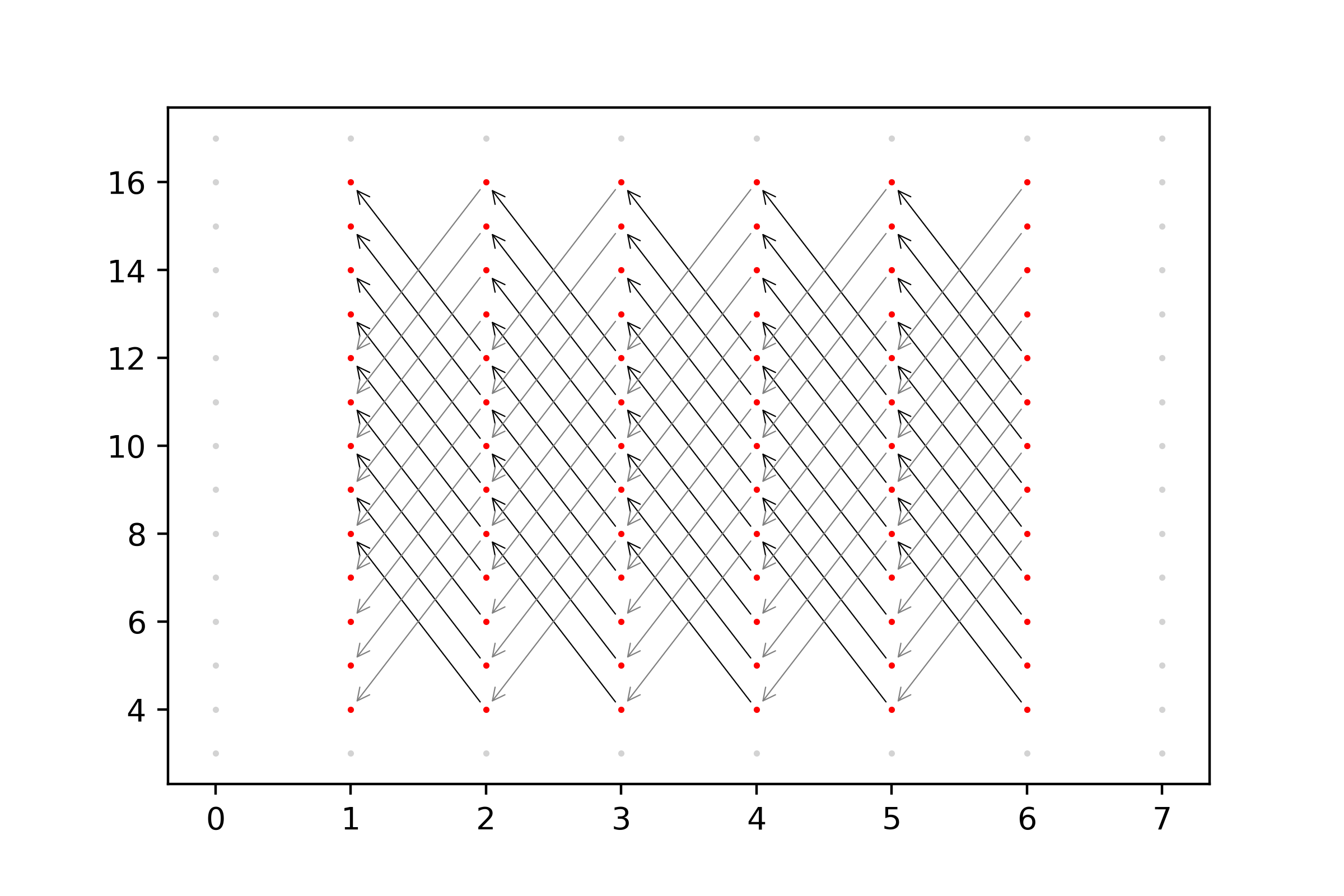}\label{fig:devito_normal}
  }
  \hfill
  \subfloat[Space-tiling, execution proceeds column-by-column]{
    \includegraphics[width=0.6\textwidth]{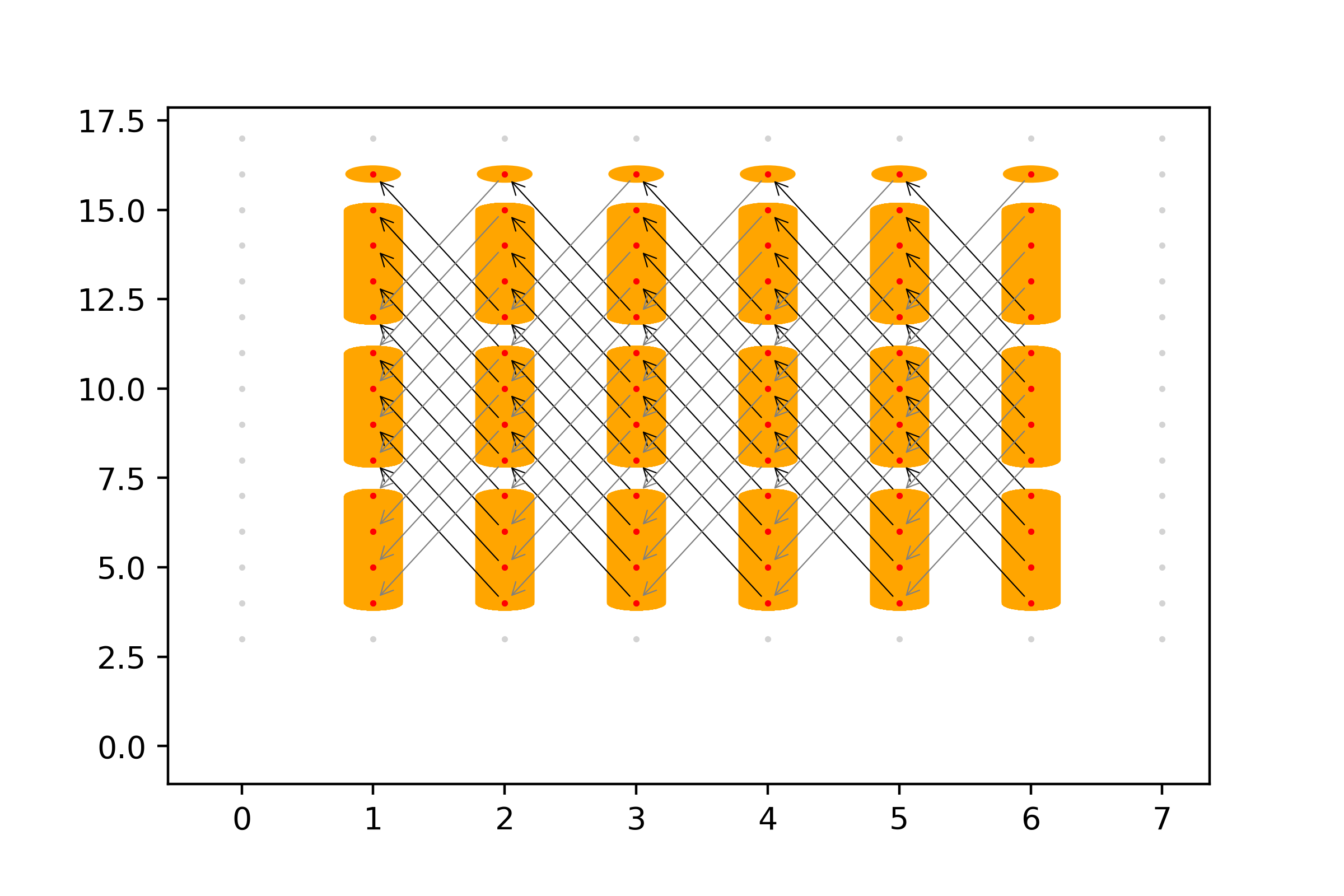}\label{fig:dep_tiled}
  }
  \hfill
  \subfloat[Time-tiling, execution proceeds column by column]{
    \includegraphics[width=0.6\textwidth]{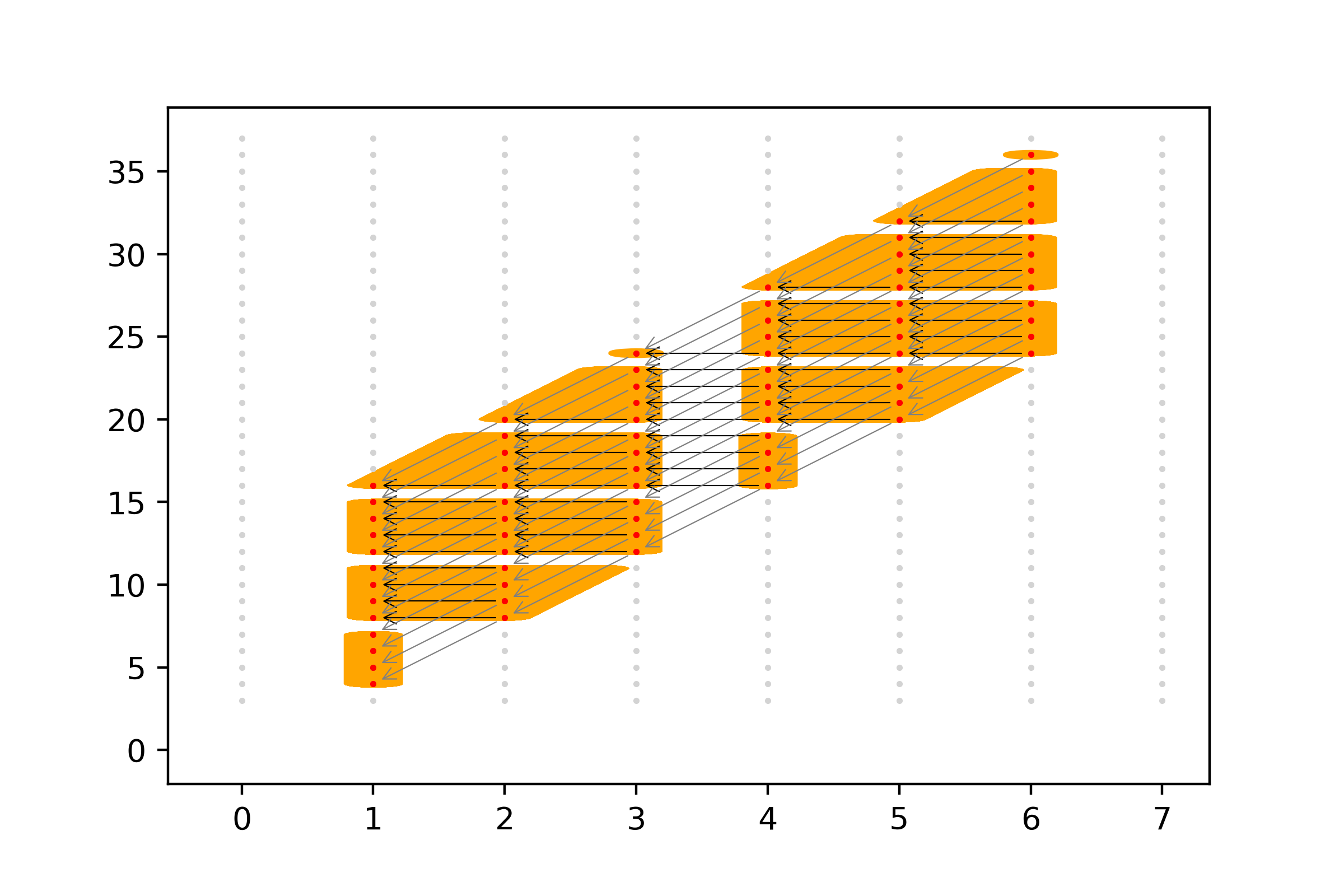}\label{fig:time_tiled}
  }
  \caption{Three iteration spaces for a stencil loop nest with dependence arrows corresponding to the 
  dependencies in \ref{fig:stencil_dependencies}. In all instances the horizontal axis is the time dimension
  and the vertical axis is the \textit{x} spatial dimension. \cite{islplot}}
  \label{fig:tiling_comparison}
\end{figure}

The iteration spaces in \ref{fig:tiling_comparison} are simplified representations of the 4D iteration
space for Devito space-tiled code and our own time-tiled code. In \ref{fig:dep_tiled} \ref{fig:time_tiled}, 
at each time step a 3D spatial tile is executed. With vectorization the third spatial dimension can be considered
as a sequence of vector statements rather than a full loop, making the spatial tiles 2D. 
Dependence arrows have been added to indicate the most `extreme' dependencies and how they change with tiling.
\ref{fig:devito_normal} shows the iteration space of a normal Devito stencil that has no tiling optimizations 
and executes each inner dimension in its entirety before moving to the next step of immediate outer dimension
as in a normal loop nest.

In \ref{fig:dep_tiled} the 1D tile shapes represent the `edge' of a 2D spatial tile (in the \textit{x} and \textit{y}
dimensions), like those shown in \ref{fig:tile_sizes} but with a tile size of 4.
In spatially-tiled code all spatial tiles are executed  before moving to the next time step. Since time
steps are completed in their entirety before moving to the next time step, all dependencies 
which point to previous time steps are respected even without any skewing.

The spatial tiling does not change in \ref{fig:time_tiled}, however the manner in which spatial tiles are executed
changes significantly with the time-tiled approach. For each time tile (delimited by the vertical `bands'),
each spatial tile (delimited by the horizontal `bands') is executed through
all the time steps within the time tile's bounds (3 in this case) before moving to the next spatial tile.
Since our AWE stencil codes update each spatial point based on surrounding points at two
previous time steps, two sequential time steps will overlap many of their memory accesses. Tiling through time allows us
to take advantage of this overlap and perform calculations for multiple time steps while the spatial data is 
still in cache memory from previous time steps. Compared to \ref{fig:skew_space}, there is significantly more
skewing in \ref{fig:time_tiled} to address the more far-reaching dependencies present in the real AWE stencil.
Again, the dependence arrows (which extend in the same manner through the omitted \textit{y} dimension) show
that this skewing is sufficient to prevent any time tile from depending on future computations.
\clearpage

\section{Transformations in CLooG}
CLooG provides two primary interfaces for describing an iteration space and the transformations
applied to it; a native C API, and a command-line interface which consumes specially formatted text files.
All of the stencils produced in this research were generated using the command-line interface \cite{cloog}
and manually produced input files \cite{my_code}. The formal structure of these input files is relatively complicated
as it is designed to support arbitrarily complex problem descriptions. However for the purposes of understanding
the input files used in this research, a simplified presentation of the input file structure will suffice.

\subsection{Input to CLooG}
\begin{figure}[h]
\begin{lstlisting}[language=C, style=customc]
for(int i = 0; i <= 10; i++){
	for(int j = i; j <= 12; j++){
    	S1(i,j)
    }
}
\end{lstlisting}
\begin{lstlisting}[language=bash, style=customsh]
# CONTEXT
# language: C
c
# Parameters & their constraints
0

# STATEMENTS
# Number of statements
1
4 4 # Number of rows, columns
#>=  i  j  1 
1   1  0  0 # i >= 0
1  -1  0 10 # i <= 10
1  -1  1  0 # i <= j
1   0 -1 12 # j <= 12
0 0 0
# we want cloog to set the iterator names
0

# SCATTERING
# Number of scattering functions
0
\end{lstlisting}
\caption{A basic \texttt{for} loop and the CLooG input file which describes it.}
\label{fig:cloog_input}
\end{figure}

Following the close relation between loop iteration spaces and more general polyhedral and
integer set mathematics, CLooG input files comprise a number of matrices that encode inequalities
from which loop domains and transformation matrices can be derived. For instance, the input files used
to produce our stencils contain two primary matrices as can be seen in \ref{fig:cloog_input}; one to specify
the `domain' or iteration space of a statement, and another to specify a `scattering' function to be applied
to the loop dimensions when generating code. Since the loop nests we are interested in tiling consist only of one
stencil computation and assignments to temporaries used in common sub-expression elimination, we 
configure CLooG to consider a single statement domain.

\subsubsection{Domain}
Each row of the input file's domain matrix represents an inequality in terms of the iterators 
(the columns of the vector) of the loop nest being considered. The ordering of the columns and the
iterators they represent is constant throughout the input file, and if there are no scattering functions
which perform scheduling or interchange transformations, also dictates the loop order in generated code.
The Devito stencil codes we transformed
contained no conditional statements within the loop nest and so the loop body has one contiguous, regular iteration
space. This means the inequalities used to describe the statement domain are effectively a translation
of the bounds of the loops being transformed into inequalities. For example, the first two rows of the matrix in 
\ref{fig:cloog_input} are inequalities representing the lower and upper bounds of the loop in \ref{fig:cloog_input}.

Loop tiling is a transformation of the iteration space which introduces new iterators and changes the bounds
of existing ones. While the overall shape and volume of the iteration space does not change after a basic tiling transformation,
the way the space is defined in terms of its iterators does. Because of this, the domain matrix in a CLooG
input file is where our tiling transformation is configured and applied. Columns are introduced for the new loops
which iterate over tiles, and inequalities are added which describe the tile loops' bounds as well as their relation
with their corresponding inner loop which iterates over points in the iteration space. For example,
consider the simple case of the single loop in \ref{fig:1d_loop}, The bounds of this loop are represented by the
inequality \[ 0 \leq i < 100\] 
or, in the polyhedral representation, the domain $D^{S1}$
\[D^{S1} = \{i \, | \, 0 \leq i < 100\}\]
To tile this loop with a tile size of 8, we would introduce the tile loop $ii$ and the inequalities representing the transformed
loop nest would become
\[0 \leq i < 100 \,\wedge\, 0 \leq ii < \floor*{\frac{100}{8}} \,\wedge\, 8*ii \leq i < 8*ii + 8  \]
or in the polyhedral representation, the domain $D_1^{S1}$
\[D_1^{S1} = \{ (ii, i) \,|\, 0 \leq i < 100 \,\wedge\, 0 \leq ii < \floor*{\frac{100}{8}} \,\wedge\, 8*ii \leq i < 8*ii + 8 \} \]
This means that for each loop being tiled, we require 4 rows (two to specify the inner loop's absolute bounds, and two to specify 
its bounds relative to the tile loop) in the domain matrix, as well as two iterator columns. 

\begin{figure}[h]
\centering
\begin{lstlisting}[language=bash, style=customsh]
  for(int i = 0; i < 100; i++){
      S1(i);
  }
\end{lstlisting}
\caption{A basic 1-D \texttt{for} loop}
\label{fig:1d_loop}
\end{figure}

Below, we give an excerpt \ref{fig:tiled_256} from one of the stencils evaluated in \autoref{sec:results} 
along with the CLooG input file \ref{fig:tile_input} used to generate its loops.

\begin{figure}[h]
\begin{lstlisting}[language=bash, style=customsh]
# language: C
c

# Parameters & their constraints
4 6
# tb ub  lb ts  1 (time upper bound, space upper and lowerbound, tile size)
1  1  0  0  0  -3
0  0  1  0  0 -276
0  0  0  1  0 -4
0  0  0  0  1 -16
1
time_size ub lb ts

# Number of statements
1
14 13
#  tt t xx  yy x  y  z tb ub lb ts  1
1 -8  1  0  0  0  0  0  0  0  0  0  0
1  8 -1  0  0  0  0  0  0  0  0  0  7
1  0  1  0  0  0  0  0  0  0  0  0 -1
1  0 -1  0  0  0  0  0  1  0  0  0 -2
# xx & x
1  0  0 -16  0  1  0  0  0  0  0  0  0
1  0  0  16  0 -1  0  0  0  0  0  1 -1
1  0 -4  0   0  1  0  0  0  0 -1  0  0
1  0  4  0   0 -1  0  0  0  1 -1  0 -1
# yy & y
1  0  0  0 -16  0  1  0  0  0  0  0  0
1  0  0  0  16  0 -1  0  0  0  0  1 -1
1  0 -4  0  0   0  1  0  0  0 -1  0  0
1  0  4  0  0   0 -1  0  0  1 -1  0 -1
# z
1  0 -4  0  0   0  0  1  0  0 -1  0  0
1  0  4  0  0   0  0 -1  0  1 -1  0 -1
0 0 0
0

# Number of scattering functions
1
7 20
# c1 c2 c3 c4 c5 c6 c7 tt  t xx yy  x  y  z tb ub lb ts  1
0  1  0  0  0  0  0  0 -1  0  0  0  0  0  0  0  0  0  0  0
0  0  1  0  0  0  0  0  0  0 -1  0  0  0  0  0  0  0  0  0
0  0  0  1  0  0  0  0  0  0  0 -1  0  0  0  0  0  0  0  0
0  0  0  0  1  0  0  0  0 -1  0  0  0  0  0  0  0  0  0  0
0  0  0  0  0  1  0  0  0  0  0  0 -1  0  0  0  0  0  0  0
0  0  0  0  0  0  1  0  0  0  0  0  0 -1  0  0  0  0  0  0
0  0  0  0  0  0  0  1  0  0  0  0  0  0 -1  0  0  0  0  0

# iterator names
1
tt xx yy time x y z
\end{lstlisting}
\caption{The CLooG input file used to generate code evaluated in \ref{fig:256_nosave_results} for tile size 16. 
To make the input files more easily
modified and understood, parameters are defined at the top of the input file and used to control grid size, default loop bounds
and tile size. To allow testing of the same tiled stencils on multiple simulation times, the time loop bound is
parameterized (as it is in non-tiled Devito stencils) on a parameter \texttt{time\_size} which is at least 3.} 
\label{fig:tile_input}
\end{figure}

\begin{figure}[h]
\begin{lstlisting}[language=C, style=customc]
for (int tt=0;tt<=floord(time_size-2,8);tt++){
 for (int xx=2*tt;xx<=min(floord(time_size+65,4),2*tt+18);xx++){
   for (int yy=max(2*tt,xx-17);
            yy<=min(min(floord(time_size+65,4),2*tt+18),xx+17);yy++){
     for (int time=max(max(max(1,8*tt),4*xx-67),4*yy-67
              time<=min(min(min(time_size-2,8*tt+7),4*xx+2),4*yy+2);time++){
        int skew = 4*time; // Skewing factor
        int t0 = (time) % 8;
        int t1 = (time + 1) % 8;
        int t2 = (time - 1) % 8;
        #pragma omp parallel
        {
          /* Flush denormal numbers to zero in hardware */
          _MM_SET_DENORMALS_ZERO_MODE(_MM_DENORMALS_ZERO_ON);
          _MM_SET_FLUSH_ZERO_MODE(_MM_FLUSH_ZERO_ON);
          #pragma omp for schedule(static)
          for (int x=max(16*xx,4*time+4);x<=min(4*time+271,16*xx+15);x++){
            for (int y=max(16*yy,4*time+4);y<=min(4*time+271,16*yy+15);y++){
              #pragma ivdep
              #pragma omp simd
              for (int z=4*time+4;z<=4*time+271;z++){
                float tcse0 = 3.04F*damp[x-skew][y-skew][z-skew];
                u[time + 1][x-skew][y-skew][z-skew] = 
                ( 
                         ( (tcse0 - 2*m[x-skew][y-skew][z-skew])* 
                                      u[time - 1][x-skew][y-skew][z-skew] )
              - 8.25142857142857e-5F*(u[time][x-skew][y-skew][z-skew - 4]
                                     +u[time][x-skew][y-skew][z-skew + 4]
                                     +u[time][x-skew][y-skew - 4][z-skew]
                                     +u[time][x-skew][y-skew + 4][z-skew]
                                     +u[time][x-skew - 4][y-skew][z-skew]
                                     +u[time][x-skew + 4][y-skew][z-skew] ) 
              + 1.17353650793651e-3F*(u[time][x-skew][y-skew][z-skew - 3]
                                     +u[time][x-skew][y-skew][z-skew + 3]
                                     +u[time][x-skew][y-skew - 3][z-skew]
                                     +u[time][x-skew][y-skew + 3][z-skew]
                                     +u[time][x-skew - 3][y-skew][z-skew]
                                     +u[time][x-skew + 3][y-skew][z-skew] ) 
                        - 9.2416e-3F*(u[time][x-skew][y-skew][z-skew - 2]
                                     +u[time][x-skew][y-skew][z-skew + 2]
                                     +u[time][x-skew][y-skew - 2][z-skew]
                                     +u[time][x-skew][y-skew + 2][z-skew]
                                     +u[time][x-skew - 2][y-skew][z-skew]
                                     +u[time][x-skew + 2][y-skew][z-skew] ) 
                       + 7.39328e-2F*(u[time][x-skew][y-skew][z-skew - 1]
                                     +u[time][x-skew][y-skew][z-skew + 1]
                                     +u[time][x-skew][y-skew - 1][z-skew]
                                     +u[time][x-skew][y-skew + 1][z-skew]
                                     +u[time][x-skew - 1][y-skew][z-skew]
                                     +u[time][x-skew + 1][y-skew][z-skew] )
                                 + (4*m[x-skew][y-skew][z-skew]* 
                                      u[time][x-skew][y-skew][z-skew] ) 
               - 3.94693333333333e-1F*u[time][x-skew][y-skew][z-skew] 
                )  / (tcse0 + 2*m[x-skew][y-skew][z-skew]);
              }
            }
        } } }
    } } } 
    \end{lstlisting}
\caption{The time-tiled loop generated by the CLooG input file in \ref{fig:tile_input}, applied to the
acoustic wave equation stencil. Formatted for clarity.}
\label{fig:tiled_256}
\end{figure}

\section{Summary}
We have given a basic definition and implementation of time-tiling as a composition of loop transformations
which prepare an iteration space so that tiling is legal and then apply tiling to all spatial dimensions and possibly
the time dimensions as well. Through the polyhedral descriptions and iteration spaces depicted, we show how time-tiling
affects the iteration space of a space-time loop and where improvements in data locality emerge. In the next chapter
we use this understanding of time-tiling and the how it can be achieved using CLooG to describe our evaluation methodology
and results for determining the run time performance improvements time-tiling offers.

\clearpage

\chapter{Evaluation}
\label{ch:evaluation}

Given the proven benefits of the spatial tiling already implemented in Devito \cite{devito-fd},
and our own understanding of how tiling affects data locality
we were confident that time-tiling (with the right configuration) should experience at least the same 
performance improvement over non-tiled code as Devito's spatial tiling. Our primary hypothesis was that
time-tiled code would run faster than the fastest spatially-tiled code produced by Devito for a given
AWE stencil simulation due to reduced memory reuse distance and, as a result, a reduction in cache misses.
Here we present our methodology for evaluating this hypothesis as well as verifying functional correctness 
of time-tiled code. We present the results obtained from a range of runtime analyses and their implications.

\section{Objectives}
Our objective for these tests was to determine if the runtime of Devito stencils can be 
improved with tiled code generated using CLooG and whether such code produces
numerical solutions that are suitably close to those produced by non-tiled stencils.

The Devito Loop Engine supports generalised loop tiling in the space dimensions so our evaluation
focused on comparing code specifically tiled with CLooG using a number of different configurations
against the existing loop tiling transformations offered by Devito as well as standard non-tiled stencils.

\section{Testing environment}
For our evaluation to be applicable to real Devito use-cases we had to ensure
that a typical Devito workflow and environment was used to generate and execute `control' code as well as 
custom-tiled code. The Devito repository contains an `examples' directory with complete
examples for symbolic problem description, code generation and data retrieval for two problem types. 
A `benchmark' Python script exists as an interface into these examples and allows for the configuration of 
different simulations 
from the command line. This benchmark script served as our primary entry point for automating different tests. 
With a small modification to the code generation pipeline we enabled the insertion of custom code from a given
file and
in this way were able to compile and execute CLooG-tiled code in a Devito environment. 
The script also allowed for automated execution and simple modification of problem parameters such as
grid dimensions, simulation time and the use of OpenMP for parallelism.\\

\begin{figure}[h]
\centering
\begin{lstlisting}[language=C, style=customc]
$ python benchmark.py run -P acoustic -dle advanced -d 512 512 512 -t 750 -so 8 -c
Applying Forward
DSE: extract_time_invariants [flops: 58, elapsed: 0.00] >>
     eliminate_inter_stencil_redundancies [flops: 58, elapsed: 0.00] >>
     eliminate_intra_stencil_redundancies [flops: 57, elapsed: 0.01] >>
     factorize [flops: 37, elapsed: 0.02] >>
     finalize [flops: 37, elapsed: 0.00]
     [Total elapsed: 0.04 s]
DLE: analyze [elapsed: 0.01] >>
     avoid_denormals [elapsed: 0.00] >>
     simdize [elapsed: 0.07] >>
     ompize [elapsed: 0.01]
     [Total elapsed: 0.08 s]
custom
IntelCompiler: compiled /tmp/devito-1000/6d6c15236570f643a716141f63e58ad716078868.c [11.65 s]
===========================================================================
loop_p_src_1<246,1> with OI=0.90 computed in 0.000s [Perf:0.27 GFlops]
loop_p_rec_2<246,101> with OI=1.94 computed in 0.001s [Perf:3.44 GFlops]
main<246,528,528,528> with OI=2.25 computed in 88.240s [Perf:15.18 GFlops]
===========================================================================
\end{lstlisting}
\caption{Shell output from running an acoustic wave simulation with a grid size of $512*512*512$ and a simulation time of 750ms using
custom code and the `advanced' mode of Devito optimizations} 
\end{figure}

The Devito framework has existing support for timing different stages of the stencil execution at the C-code level by
recording the time at loop entry and exit in C \texttt{structs}. This timing data is propagated back to the python code and so 
our modified benchmark script could retrieve these timing values and log them after each execution along with
FLOP counts calculated by Devito. The numerical output of the stencil was also gathered in the same manner for testing correctness.

\subsection{Hardware and Software}
To maximize our control over the testing environment and take advantage of some of the latest 
architectural features, our performance testing was carried in a fresh install of Ubuntu 17.04 
on a machine with a quad-core Intel i7-6700K Skylake processor running at 4.00 GHz with 16GB of DRAM. All CPU clock
scaling was disabled in the BIOS to ensure experiments always ran with the same, constant processor speed.
Additionally, no applications other than those required by the operating system were running during tests
to minimize external influence on stencil run times and memory availability.

\subsubsection{Cache hierarchy}
Since the success of tiling transformations is heavily dependent on the target architecture, we had to make sure
our Devito simulation parameters were chosen such that unoptimized code could benefit from improved data locality.
Experiments were carried out to determine which tile configurations were beneficial for the cache hierarchy of the test system.
In particular we needed to make sure that executing a single time step of a simulation required significantly more data 
than can fit in the L3 cache, which was 8MB for our test architecture. Running the acoustic wave simulation
with grid sizes of 256 and greater ensured that a single time step would need at least 64MB of data\\
\begin{table}[h]
\centering
 \begin{tabular}{||p{1.7cm} | p{2.4cm} | p{2.3cm} | p{1.5cm} | p{2.2cm} | p{2.2cm}||} 
 \hline
 Cache Level & Capacity &  Associativity & Line Size (bytes) & Fastest Latency (cycles) & Sustained Bandwidth (bytes/cycle)\\ [0.5ex] 
 \hline\hline
 L1-Data & 32KB & 8-way & 64 & 4 & ~81 \\
 \hline
 L2 & 256KB & 4-way & 64 & 12 & ~29\\
 \hline
 L3 & 8MB (2MB per core) & 16-way & 64 & 44 & ~18 \\
 \hline
\end{tabular}
\caption{Cache hierarchy information for Skylake processors \cite{intel_manual}}
\label{table:cache}
\end{table}

\subsubsection{Multithreading}
In practice the stencil codes produced by Devito are often executed in parallel using multiple threads. To keep
our performance testing realistic, we tested codes which made use of OpenMP directives to execute in parallel. Running
stencils with multithreading also created an environment where processing and memory hardware would be used to their fullest
and prevent any influence on performance by the limitations of serving a single-core application.  
In order to minimise nondeterminism and ensure even distribution of resources between threads, we used the 
Intel Thread Affinity Interface \cite{affinity} to `pin' threads to individual cores. When compiling with the Intel Compiler
\texttt{icc} \cite{icc}, the Thread Affinity Interface provides the environment variable \texttt{KMP\_AFFINITY} 
for simple configuration of thread affinity. In our tests, setting the environment variable to
\texttt{KMP\_AFFINITY=verbose,scatter} gave verbose information on thread start-up to verify our configuration, 
and the \texttt{scatter} argument ensured that different threads were allocated
to distinct cores. Since the execution of stencil codes Devito produces does not feature blocking 
or intermittent intensity and is instead constantly utilizing as many resources as it can while still being
memory-bound, hyperthreading technologies would not have been beneficial. 
With hyperthreading disabled, the i7-6700K supports 4 threads with one per core and all of our parallel
tests were executed using 4 threads.

\subsubsection{Compilation}
To take advantage of the its better understanding of Intel architecture-specific features
, such as advanced vector extensions, all evaluated code was compiled using the Intel Compiler
\texttt{icc}. The \texttt{-O3} flag was used in compilation to enable the highest optimization level
offered by \texttt{icc}, and the \texttt{-qopenmp} enabled OpenMP parallelism. We also compiled using
\texttt{icc} as it is used frequently in scientific computing applications for its advanced optimizations
and the sometimes significant improvements in run time on Intel hardware when compared to
more common compilers like \texttt{gcc}.

\section{Test parameters}
To best understand how different problem sizes reacted to different forms of time-tiling we
varied a number of simulation and code transformation parameters throughout our experiments. Below
is a description of the primary test variables and their effect on code and its execution.

\subsection{Acoustic wave equation} 
The primary simulation used for testing was the acoustic wave equation simulation for which problem set-up
and code generation code already existed in the Devito `examples' directory.
Wave equations are linked to many physical phenomena \cite{opesci-fd} and the acoustic wave equation (AWE) 
describes the propagation of acoustic waves through a medium. AWE stencils typically have a few operations 
in the main loop body which update grid points over time based on the values at other grid points that 
are nearby in both space and time. The code in \ref{fig:acoustic} is an example of the main loop kernel
for the AWE  in one dimension. The relative simplicity of this problem makes it suitable for manually-configured
experiments as  well as for understanding what transformations CLooG can perform and how to drive it to produce them.

\begin{figure}[h]
\centering
\begin{lstlisting}[language=C, style=customc]
  for(int t=1; t<T; t++){ //time loop
      for(int i=1; i<N; i++){ //1D space loop
          u[i][n+1] = 2*u[i][n] - u[i][n-1] - pow(C, 2.0)*(u[i+1][n]
                    - 2*u[i][n] + u[i-1][n]);
      }
  }
\end{lstlisting}
\caption{
    A simplified presentation of the main loop for an acoustic wave equation kernel in 1D \cite{opesci-fd}. 
    Here a point grid \texttt{u} at time \texttt{n+1} is calculated by a weighted combination of nearby points 
    \texttt{i+1} and \texttt{i-1} in the previous two time steps \texttt{n} and \texttt{n-1} \cite{opesci-fd}. }
  \label{fig:acoustic}
\end{figure}

\begin{figure}[h]
\centering
\begin{lstlisting}[language=C, style=customc]
  for (int time = 1; time < time_size - 1; time += 1) {
      for (int x = 4; x < x_size - 4; x += 1) {
        for (int y = 4; y < y_size - 4; y += 1) {
          for (int z = 4; z < z_size - 4; z += 1) {
            float tcse0 = 3.04F*damp[x][y][z];
            u[time + 1][x][y][z] = 
            	(
            	     (tcse0 - 2*m[x][y][z])*u[time - 1][x][y][z] 
                    - 8.25142857142857e-5F*(u[time][x][y][z - 4]
                                           +u[time][x][y][z + 4]
                                           +u[time][x][y - 4][z]
                                           +u[time][x][y + 4][z]
                                           +u[time][x - 4][y][z]
                                           +u[time][x + 4][y][z] ) 
                    + 1.17353650793651e-3F*(u[time][x][y][z - 3]
                                           +u[time][x][y][z + 3]
                                           +u[time][x][y - 3][z]
                                           +u[time][x][y + 3][z]
                                           +u[time][x - 3][y][z]
                                           +u[time][x + 3][y][z] ) 
                              - 9.2416e-3F*(u[time][x][y][z - 2]
                                           +u[time][x][y][z + 2]
                                           +u[time][x][y - 2][z]
                                           +u[time][x][y + 2][z]
                                           +u[time][x - 2][y][z]
                                           +u[time][x + 2][y][z] ) 
                             + 7.39328e-2F*(u[time][x][y][z - 1]
                                           +u[time][x][y][z + 1]
                                           +u[time][x][y - 1][z]
                                           +u[time][x][y + 1][z]
                                           +u[time][x - 1][y][z]
                                           +u[time][x + 1][y][z] )
                             + 4*m[x][y][z]*u[time][x][y][z]
                     - 3.94693333333333e-1F*u[time][x][y][z] 
                )/(tcse0 + 2*m[x][y][z]);
          }
        }
      }
    }
  }
  return 0;
}
\end{lstlisting}
\caption{A non-tiled Devito-generated stencil loop for the acoustic wave simulation. Non-stencil code omitted for clarity.}
\label{fig:acoustic_devito}
\end{figure}

To make our experiments realistic, the AWE stencil which we transformed and evaluated was more complicated than the
code in \ref{fig:acoustic}. In particular, our experiments considered a 4-dimensional (3-dimensional space + time) 
instance of the equation with \textit{space order} of 8 and a \textit{time order} of 2. A snippet of a non-tiled
stencil produced by Devito for this problem is given in \ref{fig:acoustic_devito}. \textit{Space order} and
\textit{time order} are simulation parameters which change the nature of the equation being solved by adjusting how 
the equation considers nearby points at previous time steps. Higher values for these parameters changes the stencil
to include points farther away in space and further back in time.

\clearpage

\subsection{Time Dimension Buffering}
Realistic simulations perform stencil computations over a large set of points for a large number of time steps, which
can require significant amounts of memory. Consider a stencil which iterates for 100 time steps
over a 600x600x600 array of floats saving the computation at each time step, the resulting data
array has size $100*600^3*4$ bytes $\approx$ 80GB. To prevent large scale simulations from requiring prohibitively large
amounts of memory, Devito supports `buffered' time dimensions, whereby only a small set of previous time steps are saved
rather than the entire history of the computation. 

For our AWE problem, Devito by default buffers 3 time steps at a time
so that regardless of the length of the simulation, only $(3*\texttt{GRID\_SIZE}^3*4)$ bytes are required. At the code-level,
this is achieved by using modular arithmetic to index into previous time steps as demonstrated in \ref{fig:tiled_256}.
Each instance of \texttt{time}, \texttt{time-1} and \texttt{time-2} in the stencil is replaced by a temporary variable which
holds the corresponding value modulo 3. Since the entirety of a time step's computations are completed before moving to 
the next time step, the modular indices are always consistent with a the moving window of three time steps required by
the stencil. Once a time step is no longer needed by a future computation it is overwritten by the new current time step,
reducing memory requirements.

Time-tiling requires the execution of a single spatial tile through multiple time steps to maximise data locality. Even
with a skewing transformation to prepare dependencies for tiling through time, the full time loop cannot be brought within
the spatial tiling loops if a buffered time dimension is used. When a tile completes execution of its time steps and a 
subsequent tiles begin execution, the buffered time array will hold data computed for the last time steps of the previous tile 
while first time steps of the new tile depend on data from the first few time steps of the previous tile. To allow time-tiling in conjunction
with time dimension buffering, the time loop is also tiled with a tile size no larger than the size of the buffered dimension
or the modulus with respect to which buffered time indices are calculated. This means each spatial tile executes one `band' of 
time steps which is small enough so that no buffered time values are overwritten before the next spatial tile begins execution,
leaving the data in the correct position for the next tile to access it using the same modular calculations.

\subsection{Simulation Grid Size}
One of the primary simulations parameters which has a large impact on memory usage and runtime is 
the simulation grid size. This parameter defines how many spatial points are in the grid being simulated 
at each time step. The grid size is related to the size of the physical area through which
acoustic wave propagation is being simulated, or the density of points in that area when considered with the parameter
which controls spacing between points. 

\subsection{Spatial Tile Size}
The dimensions of the spatial tile size impact the amount of data reuse possible within the tile and between tiles,
as well as the relative cost of executing loop control statements. Our experiments gathered performance data for a number
of spatial tile sizes ranging from 8*8 to 128*128.

\section{Code Generation}
When driven through its command-line interface, CLooG prints generated code to the command line. By running
CLooG with the optional argument \texttt{-compilable 1}, CLooG prints an entire compilable C file which includes
macro definitions for the min/max and floor/ceiling calculations it uses in loop bounds. For the results presented
here, the tiled loop nest for each stencil was generated on the command line as compilable code with CLooG and 
saved to a file before having the necessary loop structure and definitions manually inserted into a standard Devito stencil source file. To compensate for the skewing transformation, all instances of the spatial
iterators in the loop body were changed to include a subtraction of the skewing amount (the product
of the skewing factor and the time iterator), translating the
iterator values back to the original iteration space for logical correctness.

\section{Functional Correctness}
Tiling transformations can significantly change the manner in which an iteration space is traversed which in turn
can change the value of iterative computations. Before applying tiling transformations it is important to understand
the dependence structure of the code being tiled and ensure that correct ordering is preserved for statements that require it.
In the case of our Devito stencils, functionally incorrect tiling manifests itself (assuming the transformed code does not produce any
runtime errors) in the numerical output data of the tiled stencil like the \texttt{u} data array in \ref{fig:diffusion_devito}.

To simplify the challenge of tiling the acoustic wave stencil, a part of the equation which feeds data into certain grid points
at each time step had to be removed as it introduced non-uniform dependencies which could not be handled
by the time-tiling transformation we developed. Removing this `source' term from the symbolic equation and the loops it produced 
from the stencil allowed us to apply time-tiling but resulted in an empty output data array as the source equation 
gives the data array its initial values.
To compensate for this and allow for numerical verification of the tiling transformations, we modified the acoustic wave problem set-up
so that the \texttt{u} data array was initialized with a `smooth' gaussian distribution of values that made the stencil compute a measurable result.

\begin{figure}[h]
\centering
  \begin{lstlisting}[language=python, style=customc]
  def initial(dim):
            x, y, z = np.mgrid[-1.0:1.0:404j, -1.0:1.0:404j, -1.0:1.0:404j]
            xyz = np.column_stack([x.flat, y.flat, z.flat])
            mu = np.array([0.0, 0.0, 0.0])
            sigma = np.array([0.40, 0.40, 0.40])
            covariance = np.diag(sigma**2)
            gc.collect()
            x = multivariate_normal.pdf(xyz, mean=mu, cov=covariance)
            x = x.reshape((404, 404, 404))
            return x
 \end{lstlisting}
\caption{This code was added in the problem set-up phase to initialize the first time step of \texttt{u} data array
with a gaussian distribution. In this case the array is for a grid of size 384*384*384, but with `ghost cells' surrounding
grid points, the data array is of size 404*404*404. \texttt{multivariate\_normal.pdf} is a function provided
by SciPy for calculating multivariate distributions. The standard deviations were chosen so that three standard deviations
from the centre the value fell to near zero.}
\label{fig:guassian_init}
\end{figure}

\begin{figure}[h]
\centering
  \begin{lstlisting}[language=python, style=customc]
  # Simulate using custom code located in /tmp/devito-1000/custom.
  _, _, _, [customrec, customdata] = acoustic_examples.run(custom=True, **parameters)
  # Simulate with the same problem parameters, but using Devito code
  _, _, _, [rec, data] = acoustic_examples.run(custom=False, **parameters)
  # Compare data arrays
  assert np.allclose(data, customdata, atol=10e-6, equal_nan=True)
  # Compare receiver arrays
  assert np.allclose(rec, customdata, atol=10e-6, equal_nan=True)
\end{lstlisting}
\caption{A snippet from the benchmark script which runs two distinct acoustic wave simulations and compares
		 the resulting data using numpy} 
\label{fig:functest}
\end{figure}

The NumPy library provides a \texttt{numpy.allclose} function for comparing the elements of two NumPy arrays. 
For verifying the numerical output of tiled code, an absolute tolerance of $10^{-6}$ was chosen in line with 
Devito's own unit tests. Tiled code was deemed functionally incorrect if the `data' output array
did not satisfy this tolerance and produced a \texttt{False} return value from \texttt{numpy.allclose}
The code in  \ref{fig:functest} was added as additional operation mode in the benchmark script so 
that test parameters could  easily be kept consistent between performance and functional testing.

\section{Results}
\label{sec:results}
All evaluated time-tiled codes can be found at \cite{my_code}. In this section we visualize and summarise
run times and GFLOPS recordings gathered by repeatedly executing different configurations of the AWE simulation. 
All run time and GFLOPS figures were calculated with Devito's own timing mechanisms and represent only the performance 
of the main stencil loop to which our transformations have been applied.
\subsection{Transformation Types}
For each of the experiment results presented below, the following names are given to the transformations tested;
\begin{itemize}
\item \textbf{Devito-generated control stencils}
  \begin{itemize}
    \item \texttt{normal}: A standard Devito-generated stencil with only SIMD vectorization and
                            OMP parallelism optimizations enabled.
    \item \texttt{blocked\_auto}: A stencil generated with the same configuration as \textit{normal} but also with
    						the auto-tuned blocking optimization enabled. 
  \end{itemize}
\item \textbf{Time-tiled stencils}: Each of these stencils was manually transformed using code generated by CLooG specifically 
									 for each spatial tile size and simulation grid size. 
                                     Only the $x$ and $y$ spatial loops are tiled in these stencils, the $z$ loop is kept as a 
                                     full loop 
                                     to match the approach taken by Devito's own cache blocking optimization and to take better
                                     advantage of vectorization opportunities by having a higher trip count for the innermost loop.
                                     Stencil \texttt{time\_tiled\_X} has spatial tile dimensions of \texttt{x}*\texttt{x} 
                                     in $x$ and $y$.
    \begin{itemize}
      \item \texttt{time\_tiled\_8}
      \item \texttt{time\_tiled\_16}
      \item \texttt{time\_tiled\_32}
      \item \texttt{time\_tiled\_128}
    \end{itemize}
\end{itemize}

We present results for stencils generated with and without time dimension buffering;
  \begin{itemize}
  \item \textbf{Full time dimensions}: Every time step of these simulations was saved in the data array and so no
  		modulo computations were used for indexing in time allowing the time loop to be directly moved within
        the spatial tile loops to exploit reuse in the time dimension.
  \item \textbf{Buffered time dimensions}: The following stencils used a buffered time dimension of size 8 and 
  		the time loops were tiled accordingly with a tile size of 8. 
  \end{itemize}

\clearpage
\subsection{Grid size 256 with full time dimension}

\begin{figure}[h]
\centering
\centerline{\includegraphics[scale=0.60]{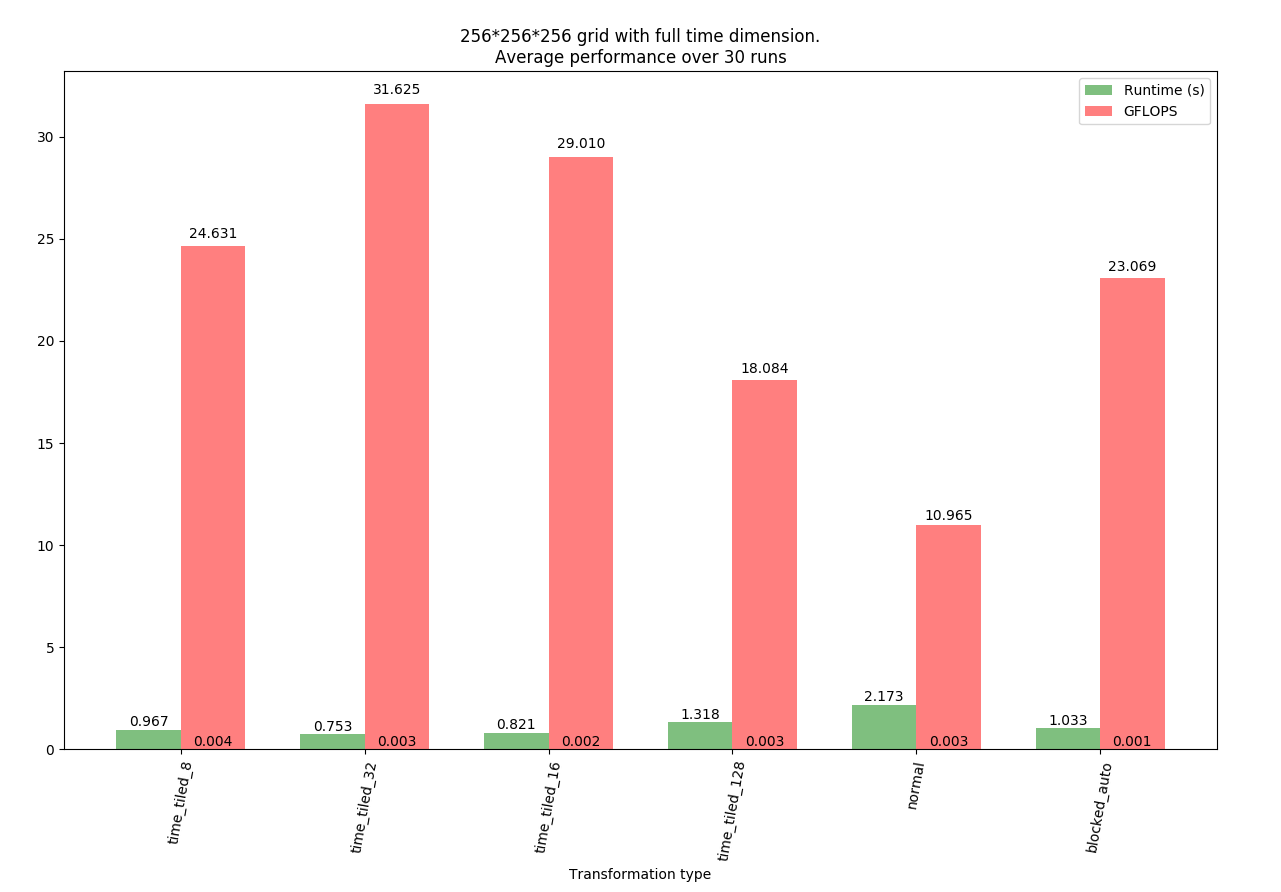}}
\caption{Runtime (green, lower is better) and GFLOPS (red, higher is better) recordings for 
Devito-generated and time-tiled stencils. Each measurement is an average over 30 runs of 
the Acoustic Wave Equation simulation, on a $256*256*256$ grid with a simulation time of 100ms. 
At the bottom of the GFLOPS bars is the standard deviation of the results.
For this problem the auto-tuned block dimensions are $128*128$ in $x$ and $y$ (\texttt{blocked\_auto}).
Each \texttt{time\_tiled\_X} result is for a time-tiled stencil with a spatial tile size of \texttt{X}*\texttt{X}.
Time-tiling with a spatial tile size of $32*32$ gave the lowest run times for this simulation, with a 27\% improvement
over Devito's auto-tuned blocking.}
\label{fig:256_save_results}
\end{figure}

In this experiment no time dimension buffering was used, meaning every time step of the computation was saved
in the data array. This put a limit on both the grid size that could be used and the length of the simulation time.
To have a realistic grid size and a simulation time long enough so that run time would not be 
too small to draw conclusions from but operate within the memory limits of our test architecture,
a 100ms simulation time was chosen on a grid size of $256*256*256$.

Time-tiling with a spatial tile size of $32*32$ gave the lowest run times for this simulation, with a 27.1\% improvement
over Devito's auto-tuned blocking. Only the time-tiled code with spatial tiles of size $128*128$ produced a slower
run time than the auto-tuned blocking result despite the fact the two transformations used the same spatial tile size. 
This could be because the increased spatial tile size results in data being evicting data from the cache during a time 
step, preventing reuse along the time dimension.
\clearpage

\subsection{Grid size 256 with time dimension buffering}
\begin{figure}[h]
\centering
\centerline{\includegraphics[scale=0.60]{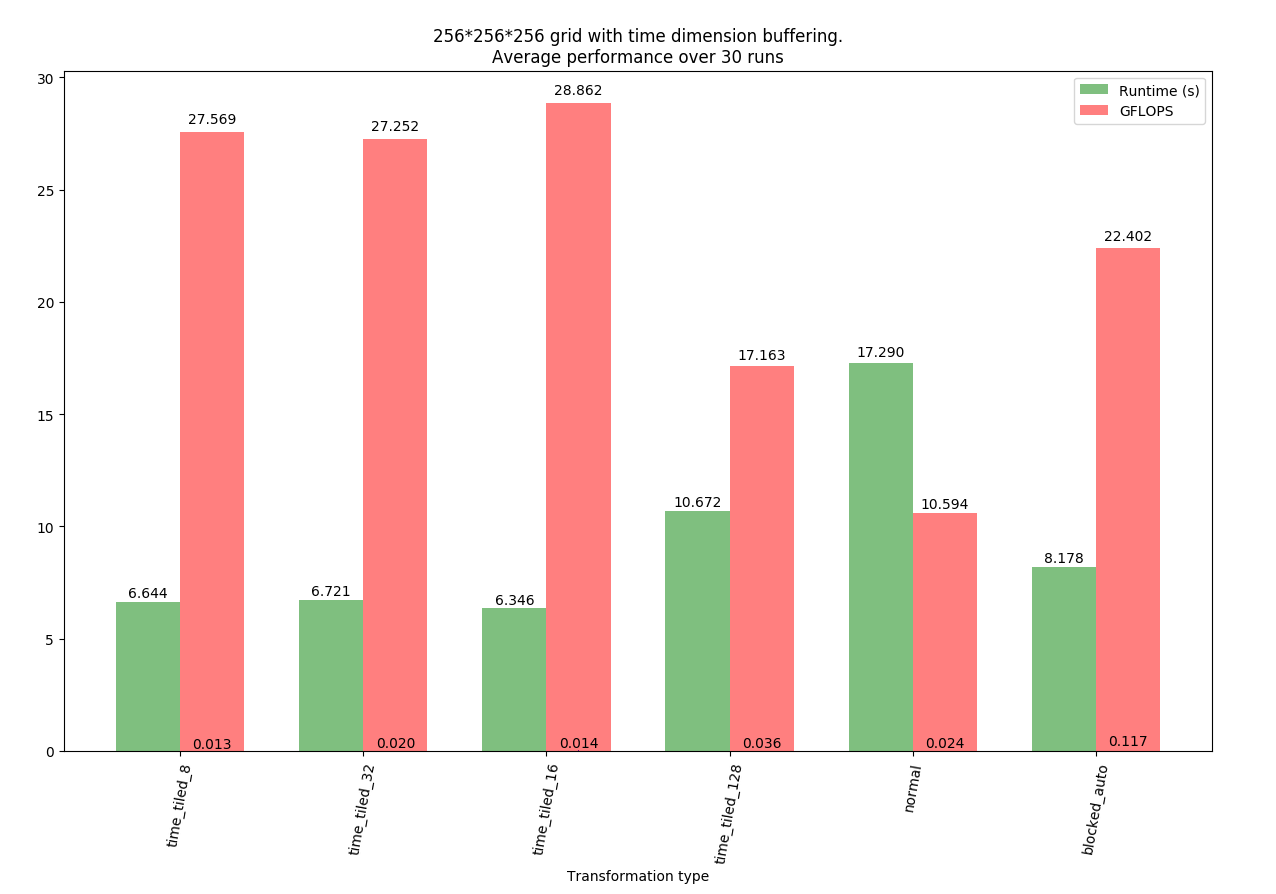}}
\caption{Runtime (green, lower is better) and GFLOPS (red, higher is better) recordings for  Devito-generated 
and time-tiled stencils. Each measurement is an
average over 30 runs of the Acoustic Wave Equation simulation, on a $256*256*256$ grid with a simulation time of 750ms. 
The time dimensions are buffered with a buffer size of 8. At the bottom of the GFLOPS bars is the standard deviation of the results.
For this problem the auto-tuned block dimensions are $32*32$ in $x$ and $y$ (\texttt{blocked\_auto}).
Each \texttt{time\_tiled\_X} result is for a time-tiled stencil with a spatial tile size of \texttt{X}*\texttt{X}.
Time-tiled code using spatial tile sizes of $16*16$ produced the lowest run time with a 22.4\% improvement over
Devito's cache blocking.}
\label{fig:256_nosave_results}
\end{figure}

\begin{figure}[h]
\centering
\centerline{\includegraphics[scale=0.60]{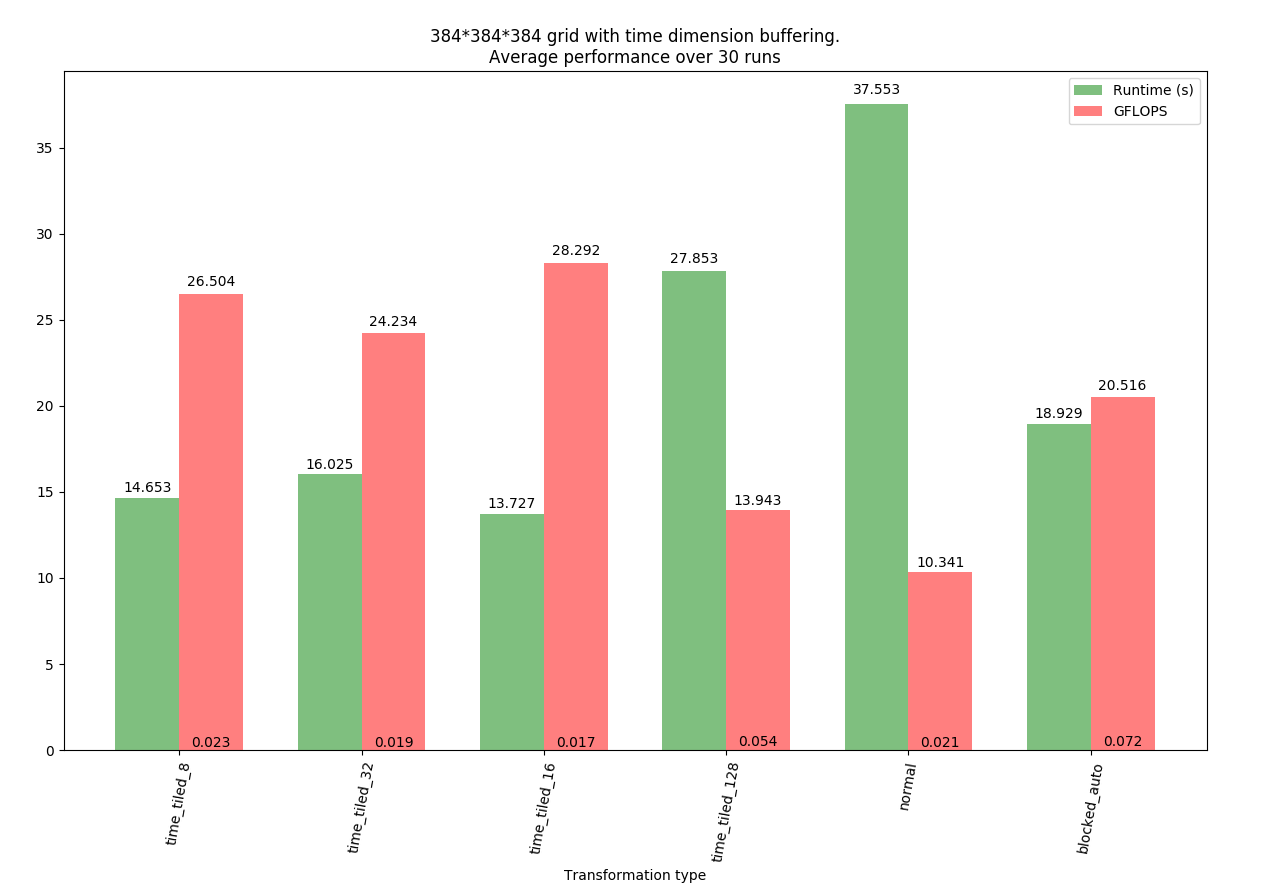}}
\caption{Runtime (green, lower is better) and GFLOPS (red, higher is better) recordings for  Devito-generated 
time-tiled stencils. Each measurement is an
average over 30 runs of the Acoustic Wave Equation simulation, on a $384*384*384$ grid with a simulation time of 750ms. 
The time dimensions are buffered with a buffer size of 8. At the bottom of the GFLOPS bars is the standard deviation of the results.
For this problem the auto-tuned block dimensions are $32*32$ in $x$ and $y$ (\texttt{blocked\_auto}).
Each \texttt{time\_tiled\_X} result is for a time-tiled stencil with a spatial tile size of \texttt{X}*\texttt{X}.
Time-tiled code using spatial tile sizes of $16*16$ produced the lowest run time with a 27.5\% improvement over
Devito's cache blocking.}
\label{fig:384_nosave_results}
\end{figure}

With time dimension buffering enabled, we were free to increase the simulation time without being limited by memory
capacity. To keep simulation time somewhat realistic and aid comparison with the results in \ref{fig:384_nosave_results}
a simulation time of 750ms was chosen.

Time-tiled code using spatial tile sizes of $16*16$ produced the lowest run time with a 22.4\% improvement over
Devito's cache blocking.
\clearpage

\subsection{Grid size 384 with time dimension buffering}
To investigate how time-tiling performed with larger more realistic spatial grids but still remain within the
memory bounds for testing functional correctness, we chose a grid size of $384*384*384$ which was close
to the maximum grid size we could test.
Time-tiled code using spatial tile sizes of $16*16$ produced the lowest run time with a 27.5\% improvement over
Devito's cache blocking.
\clearpage

\subsection{Roofline Model}
The roofline model is used to express how a program with given arithmetic intensity
performs relative to the theoretical maximum limits imposed by a machine's computational resources
and memory bandwidth. It is typically a graphical representation of these figures and limits
which demonstrates whether a computation is memory-bound or compute-bound.
For simplicity, we take some of the calculations and comparisons of the roofline model and present
them in a tabular format below. 

The AWE stencil is a memory-bound computation and this is shown by a simple calculation
of peak stencil GFLOPS. Using the STREAM \cite{stream} benchmark we calculated the peak memory bandwidth
of our test architecture to be 15.168 GB/s. Devito calculates a stencil's arithmetic intensity
(the number of floating point operations required per byte of memory traffic)
from its knowledge of the grid dimensions and the calculations performed by stencil. The peak
FLOP count for our test architecture can be calculated as the product of FLOPS per cycle
and clock speed (cycles per second). The Skylake architecture can execute 32 single-precision
floating point operations per cycle using fused-multiply-add extensions \cite{intel_manual}, which at a clock speed
of 4.00GHz gives us a peak performance of 128 GFLOPS. It is clear that the peak stencil GFLOPS presented 
in \ref{table:maxperformance} are significantly lower than the peak GFLOPS of our test architecture.
This means that our peak memory bandwidth prevents the AWE stencil it from reading data 
fast enough to ever use more than about 25\% of the floating point resources that are available, making the
computation memory-bound.

We compare our actual performance results from \ref{fig:256_save_results}, 
\ref{fig:256_nosave_results}, \ref{fig:384_nosave_results}
with the calculated theoretical maxima to get an understanding for how much more performance
can be obtained through optimizing the stencil's memory usage. The actual floating point performance for our 
test architecture shouldn't have an effect on the GFLOPS figure for a given stencil because the 
computation is memory-bound and so we are never operating at or near the peak architecture GFLOPS.
However peak memory bandwidth does fluctuate during execution and will affect the actual performance of 
our stencils by changing the rate at which they can transfer data. Likely the largest cause for deviation
from the calculated maxima however, is cache-misses due to poor data reuse. When the stencil
attempts to access memory which it has already accessed but fails to find it in cache memory, it not only
incurs a performance penalty as it waits for the data to be read from main memory, but it also increases
the total memory traffic from the theoretical minimum. These effects can be seen in the \textit{\textbf{Change}}
columns of \ref{table:roofline}. The data suggest that for despite the run time improvements achieved with time-tiling,
there is still remove for improving data reuse in long-running simulations that make use of time dimension buffering.

\newcommand{\rr}{\raggedright}
\begin{table}[h]
\centering
 \begin{tabular}{||p{1.0cm} | p{1.6cm} | p{1.0cm} | p{1.9cm} | p{1.6cm} | p{2.0cm} | p{1.2cm} ||} 
 \hline
 Grid size & Time buffering & Sim. time (ms) & Arithmetic Intensity & Stencil GFLOPS & 
 \rr Memory traffic (GB) & Run time (s) \\ [0.5ex] 
 \hline\hline
 256 & No  & 100 & 2.15 & 32.612 & 11.101 & 0.732 \\ \hline
 256 & Yes & 750 & 2.20 & 33.37  & 83.089 & 5.478 \\ \hline
 384 & Yes & 500 & 2.25 & 34.129 & 174.079 & 11.477 \\ \hline
\end{tabular}
\caption{Theoretical minimum run time calculations for the stencils evaluated in \ref{sec:results}. 
Arithmetic intensity (number of floating point operations per byte of memory traffic) and
minimum memory traffic figures are derived by Devito from its knowledge of the stencil computation.
We calculate peak GFLOPS for a given stencil as the product of its arithmetic intensity and
the peak bandwidth of our architecture (measured as 15.168GB/S using the STREAM benchmark \cite{stream}.
Theoretical minimum run time is calculated as the quotient of memory traffic and memory bandwidth.
In practice, memory bandwidth and GFLOPS will not always be at their peak.}
\label{table:maxperformance}
\end{table}

\begin{table}[h]
\centering
 \begin{tabular}{||p{0.9cm} | p{1.6cm} | p{1.0cm} | p{1.55cm} | p{1.55cm} || p{1.65cm} || p{1.6cm} | p{1.3cm} || p{1.65cm} ||} 
 \hline
 Grid size & Time buffering & Sim. time (ms) & Stencil GFLOPS & 
 Actual GFLOPS & \textbf{Change} & Minimum run time (s) & Actual run time (s) & \textbf{Change} \\ [0.5ex] 
 \hline\hline
 256 & No  & 100 & 32.612 & 31.625 & \textbf{-2.03}\%  & 0.732  & 0.753  & \textbf{-2.87}\%  \\ \hline
 256 & Yes & 750 & 33.37  & 28.862 & \textbf{-13.51}\% & 5.478  & 6.346  & \textbf{-15.84}\% \\ \hline
 384 & Yes & 500 & 34.129 & 28.929 & \textbf{-15.24}\% & 11.477 & 13.727 & \textbf{-19.60}\% \\ \hline
\end{tabular}
\caption{A comparison of peak theoretical performance figures calculated in \ref{table:maxperformance}
and actual performance results (the same average results plotted in \ref{sec:results}).
By comparing our actual performance to the maximum calculated performance, we get an understanding
of how much room for optimization remains. Suboptimal data reuse in our stencils results in increased
actual memory traffic for the stencils, which when combined with fluctuating memory bandwidth
results in the deviations highlighted in the \textbf{Change} columns. The figures suggest there
is room for improving the data locality of time-tiled stencils with long
simulation times and time dimension buffering.}
\label{table:roofline}
\end{table}
\clearpage

\section{Future experiments}
As well as the experiments presented above, we evaluated the AWE simulation on larger grid sizes (up to $512*512*512$) 
but reached memory limits during the numerical verification phase, making the data gathered unreliable.
Attempting to run Devito's auto-tuned cache blocking optimization without any custom-tiled code was also restricted
by memory resources  due to the increased overall memory requirements by augmenting the time dimension to support tiling 
through time (typically a buffered time dimension has size 3, but to support tiling we enlarged the dimension to size 8).
In addition to performance testing, memory analyses should be carried out to measure cache misses, cache miss rate,
and total memory traffic to make certain that run time performance improvements result from better cache performance
and data reuse.

\section{Conclusion}
In this chapter we have shown that for a simplified version of the acoustic wave equation stencil
generated by Devito, time-tiling can offer significant improvements in run time. We
evaluated time-tiling transformations on three different simulations, varying grid size and the use
of time dimension buffering and saw improvements with time-tiling in every case. The simulation
with the largest grid size (384*384*384) experienced the most significant speed-up due to time tiling.
The results presented here are not enough to determine whether run time improvements are due to 
fewer cache misses and reduced data reuse distance, and so our hypothesis has been partially
proven. However the data are enough to warrant further investigation with different simulations 
and tiling parameters.

\chapter{Conclusion}
\label{ch:conclusion}

To conclude, we summarise the contributions of this research and its supporting evidence.
Using the insight gained through this investigation, we comment on potential designs for automating
time-tiling in Devito and what future work can be done to strengthen our claims and 
apply time-tiling to a broader range of stencil codes.

\section{Summary of Contributions}
This report documents our approach to investigating the nature and feasibility of time-tiling transformations
which can improve the run time performance of space-time stencil loops within the Devito finite-difference
code generation framework. We motivated our research with the fundamentals of code optimization within the polyhedral model,
before presenting and evaluating a basic time-tiling transformation. We now summarise the main contributions of this project;
\begin{itemize}
\item We identify opportunities for tiling through the time dimension of Devito stencil loops to improve data reuse. Devito
currently supports tiling in the spatial dimensions of its stencil loops, but by the nature of space-time stencil
computations there is an often significant overlap of memory accesses between subsequent time steps which is not exploited
by spatial tiling. 
By executing multiple time steps for each spatial tile, we benefit from improved cache performance by keeping spatial data in
cache memory from one time step to the next so that it may be reused.
\item We give conditions which define the legality of a time-tiling transformation. A Devito stencil's iteration space
cannot contain dependencies with positive dependence distances in any of the spatial dimensions.
\item We propose time-tiling as a composition of loop transformations which prepare a stencil's iteration space
so that the time loop can be moved within the spatial tile loops. Our research in the context of uniform dependencies
like those seen in the acoustic wave equation stencil find that time-tiling can be achieved by skewing the each spatial dimension
according to the greatest positive dependence distance in that dimension, and then applying spatial tiling.
\item We show that time-tiling can still be applied to Devito loops which make use of time-dimension buffering to
reduce overall memory requirements. By tiling the time dimension in addition to the space dimensions, and ensuring the size
of the tile in the time dimension is less than or equal to the modulus with respect to which time dimension buffering calculations
are performed.
\item We give our methodology for verifying the numerical correctness of time-tiled Devito stencils to validate
our performance experiments.
\item We evaluate the performance of time-tiling using the multi-threaded Devito acoustic wave equation stencil. 
Our experiments tested 
several simulation parameters for the AWE stencil, namely grid size, simulation time and full/buffered time dimensions.
Multiple spatial tile dimensions were also tested to find out which spatial tile sizes offered the best run time improvements
and how they compared to the best performing spatial tile sizes chosen in Devito's own cache-blocking optimization.
\item We find that time-tiling can offer run time improvements of up to 25.7\% when applied to the AWE stencil with
relatively small grid sizes and simulation times.
\item We demonstrate that compared to theoretical maximum performance figures, our time-tiled stencils can still benefit from
further improved data reuse and reduced overall memory traffic.
\item We give our testing methodology to demonstrate how we ensured the reliability of our performance results.
\end{itemize}

\begin{table}[h]
\centering
 \begin{tabular}{||p{1.9cm} | p{2.4cm} | p{2.4cm} | p{2.4cm} | p{2.3cm} | p{2.3cm} ||} 
 \hline
 Grid size & Time dimension buffering & Auto-tuned blocking (s) &  Time-tiling (s) & Improvement & Spatial tile size\\ [0.5ex] 
 \hline\hline
 256 & No & 1.033 & 0.753 & 27.1\% & 16\\
 \hline
 256 & Yes & 8.178 & 6.346 & 22.4\% & 32\\
 \hline
 384 & Yes & 18.929 & 13.727 & 27.5\% & 32\\
 \hline
\end{tabular}
\caption{Runtime results for the best performing time-tiled stencils in each evaluated simulation. `Spatial tile size'
refers to the size of the \textit{x} and \textit{y} dimensions of the spatial tiles in the time-tiled code. `Auto-tuned blocking'
results are the run times of Devito's own auto-tuned spatial tiling transformation. All data presented here is taken from
\ref{fig:256_save_results}, \ref{fig:256_nosave_results}, \ref{fig:384_nosave_results}.}
\label{table:results}
\end{table}

\section{Future Work}
In acknowledgement of the shortcomings of this research, we propose areas where we feel future work would be particularly
profitable.
\begin{itemize}
\item Carry out further evaluation of time-tiled AWE stencils produced with different tiling and simulation parameters.
\item Investigate the statement and dependence structure of `source' term injection loops to determine how to tile them.
\item Develop the fundamental time-tiling transformations presented in this research to apply to more advanced Devito stencils.
\item Automating general-case time-tiling transformations in Devito, possibly using the suggestions made in \autoref{subsec:automation}.
\end{itemize}

\subsection{Automating time-tiling in Devito}
\label{subsec:automation}
The transformations and results presented here serve as a proof-of-concept for time-tiling Devito stencils. However,
for users to experience the claimed performance improvements when running more realistic simulations on arbitrary
stencil problems, a robust system must be developed which uses the analytical and code generation power
of Devito to automatically generate time-tiled stencil loops. In this section we discuss some of the available tools and possible
designs that could be used to implement an automatic time-tiling optimization pass in Devito. 

\subsubsection{Tools}
During this investigation, we researched several currently available tools for code generation in
the polyhedral model. Each take different approaches to generating code (i.e loops) which scans the points of convex polyhedra
and also provide different interfaces and outputs, affecting their suitability for use in automated time-tiling of Devito loops. 
\begin{itemize}
\item \textbf{CLooG}:
CLooG (\autoref{sec:cloog}) uses the  Quillere et al. algorithm \cite{polyhedra-algo} to generate loops
which iterate over every point in a given union of convex polyhedra (statement domains) according to a given schedule.
It provides a command-line interface \cite{cloog} for producing code according to a specification provided as a specially-formatted 
text file and offers a number of command line options for controlling the size, complexity and syntax of the generated code.
Additionally, CLooG can be used as a C library which gives finer control over CLooG's options and facilitates 
programmatic description of arbitrary domains and schedules. Using CLooG as a library also gives direct access to
generated code as an AST, which is preferable for further code generation analysis as it avoids working with strings.
Being purpose-built for code generation, CLooG is the most attractive option for tiling code generation in Devito
when not considering logistical details.

\item \textbf{isl}:
The integer set library (\texttt{isl}) is a C library for describing and manipulating sets of integer points
bounded by affine constraints which is primarily developed by Sven Verdoolaege \cite{isl}. 
It is still in development, but \texttt{isl} provides a number of powerful features
including dependence analysis, affine scheduling of iteration domains, and AST generation. CLooG uses \texttt{isl}
as a polyhedral back-end for manipulating iteration spaces and determining how best to scan them. \texttt{isl} is a
comprehensive library with thousands of functions so it is not designed to be a code-generation tool, though it
can act as one. The library can be seen as the generalized implementation which a number of other polyhedral 
tools attempt to specialize and improve on. \texttt{isl} comes with extensive documentation.

\item \textbf{islpy}:
\texttt{islpy} is a set of largely auto-generated Python bindings for \texttt{isl}, and is developed by Andreas 
Kloeckner \cite{islpy}. As a wrapper around \texttt{isl}, \texttt{islpy} allows developers to enjoy the naturally
expressive syntax of Python along with the power tructures of \texttt{isl} in C.
\end{itemize}

\subsubsection{Challenges}
Using these tools and developing an automated workflow for producing time-tiled stencil loops in Devito
faces a number of challenges technical and conceptual in nature.
\begin{itemize}
\item \textbf{CLooG}: This report demonstrates clearly
that CLooG is capable of generating high-performance time-tiled code for at least Devito's acoustic wave equation stencil,
however we have only demonstrated how to drive CLooG towards this using its command line interface and text-based input files.
For efficient, robust interaction between Python-based Devito and CLI/C-based CLooG cannot rely on the creation and parsing of
textual problem descriptions and output code, particularly for more complex Devito stencils. CLooG's API is a more natural choice
since it can produce abstract syntax trees, however there is some significant communication infrastructure which would need
to be implemented for communicating between Devito and the C API, and translating CLooG's AST output in C to the AST format
used by Devito. The inclusion of CLooG would also create additional dependencies for the Devito project and could potentially
complicate the build process/requirements for end users. The PLuTo project (which is largely written in C) shows that
CLooG can be used as a code-generation back end for advanced loop optimizations, but makes use of different technologies than
Devito.

\item \textbf{isl and islpy}: Being a C library, \texttt{isl} faces the same interfacing problems as CLooG when considered
for automating time-tiling in Devito. \textit{isl} is not designed primarily as a code generation tool and so may not
provide some of the advanced configurations or algorithms employed by CLooG. The size and complexity of \textit{isl} 
also poses a significant challenge for a developer looking to use and understand its API. The python wrapper \textit{islpy}
offers a solution to the challenges of interfacing between Devito and isl, but would still be challenging during implementation.
Our own experiments with \texttt{isl} following the few published examples for code generation were successful but only when using
the `helper' functions provided to simplify construction of domains and schedules using string representation, which is not 
suitable in an automated optimization context. If the initial challenge of developing a time-tiling workflow using 
\texttt{isl} functions could be overcome, \texttt{islpy} seems like the most suitable of the described tools for
time-tiling Devito stencils.

\item \textbf{Depedencies}:
As well as the technical difficulties of Devito delegating to a polyhedral tool for time-tiling,
it is possible that with more advanced applications, data dependencies may pose a greater challenge than they
do in the stencils we have evaluated.
Our research has made clear the importance of ensuring a time-tiling is legal with respect to a
stencil's data dependencies. The stencils evaluated in this report featured relatively simple, uniform dependencies
which were the same in all spatial dimensions. More complex stencils often feature similar dependence structures
with increased arithmetic load, however for loops with non-uniform dependencies such as those featured
in Devito `source' and `receiver' loops it may be difficult to determine what iteration space transformations
are required for time-tiling to be legal. For time-tiling to be applicable to realistic Devito stencils, a
solution to problem of tiling the `source' and `receiver' loops must be found.

\item \textbf{Cost analysis}:
The acoustic wave equation stencil considered in our research has its performance bounded by the memory
bandwidth of the architecture it is being run on. It has a relatively low arithmetic intensity meaning
that more load is placed on memory transfer resources than on arithmetic units, thus creating an opportunity
for the stencil's run time to be decreased by transformations which improve data locality (such as tiling).
However other stencil have much higher arithmetic intensities, meaning their run time is bounded by the amount
of arithmetic resources available rather than memory bandwidth. In this case tiling transformations may not be very 
profitable and could be detrimental to stencil performance, highlighting the need for some heuristic or process by which
Devito determines if time-tiling transformations are worthwhile. This may not be very challenging given Devito's
understanding of the arithmetic and memory requirements for a given stencil and a user's understanding of their hardware
limitations.
\end{itemize}

\subsubsection{Implementation}
Putting aside the technical details of communication between Devito and a polyhedral tool used for generating
time-tiled loops, we outline a possible high-level pathway for implementing automated time-tiling in Devito. Before
beginning implementation of an time-tiling optimizing pass, it is important to carry out further experimental work
to ensure the promising run time improvements presented in this research can be reproduced in more advanced
Devito stencils.
\begin{enumerate}
\item Investigate whether time-tiling transformations are feasible for different types of stencil code and judge whether
time-tiling is worth automating for the general case.
\item Implement an optimizing pass using the Devito Loop Engine which retrieves the following (readily available) information 
from Devito's AST representation that can be used to drive a code-generation tool; 
\begin{itemize}
\item Whether a loop should be vectorized and thus not tiled. 
\item Explicit bounds for all stencil loops being tiled to prevent code explosion which can occur when using parameterized polyhedra.
\item For more complex loops it may be necessary to identify or derive conditionals which partition their iteration space
		and produce a representation of these conditionals in terms of the loop iterators so that they can be communicated
        to the code-generation tool.
\item The dependence structure of the stencil. In particular, for uniform dependencies like those in the AWE stencil,
		we require the dependencies which have `the most negative' dependence distance in each dimension. For time-tiling
        to be legal, each dimension should be skewed by the corresponding distance so that all dependencies in the tiled
        domain are lexicographically positive.
\item Whether time dimension buffering is enabled.
\item A representation of iterator names or identifiers for tracking the tiling parameters and stencil information pertaining
to each loop
\end{itemize}
\item Depending on the tiling configurations specified by the user or Devito's settings, additional information should be attached
to each domain which specifies the tile size and skewing factor required.
\item This data can be passed from the optimizing pass to an interfacing component which uses it to prepare domains and schedules in the 
accepted format of a polyhedral code-generation tool according to the tiling parameters and the information obtained from the AST. This component
delegates to the tool to generate a time-tiled AST and handles parsing of input/output before returning the transformed AST
to the DLE.
\item The time-tiling pass should augment the any references to loop iterators in the stencil body
to correct for any skewing applied to their loop domains. An example of this can be seen in \ref{fig:tiled_256} and the subtraction of \texttt{skew} from all loop indices in the body.
\item The optimizing pass updates its AST with the new, time-tiled structure.
\end{enumerate}


\addcontentsline{toc}{chapter}{Bibliography}

\bibliography{bibliography} 
\bibliographystyle{ieeetr}
\end{document}